\documentclass[Transsize,journal]{IEEEtran}
\usepackage{amsmath,amsfonts}
\usepackage{algpseudocode}
\usepackage{algorithm}
\usepackage{array}
\usepackage[caption=false,font=normalsize,labelfont=sf,textfont=sf]{subfig}
\usepackage{textcomp}
\usepackage{stfloats}
\usepackage{url}
\usepackage{verbatim}
\usepackage{graphicx}
\usepackage{cite}
\usepackage{amssymb}
\usepackage{bm}
\usepackage{makecell} 
\usepackage{soul, color, xcolor}

\newcommand{\citerange}[2]{[\citenum{#1}]-[\citenum{#2}]}

\begin{document}

\title{Robust Beamforming Design for STAR-RIS Aided RSMA Network with Hardware Impairments}

\author{Ziyue Wang, Xiaoyan Ma, Xingyu Peng, Zheao Li, Jinyuan Liu, Yongliang Guan,~\IEEEmembership{Senior Member, IEEE}, \\and Chau Yuen,~\IEEEmembership{Fellow, IEEE}

\thanks{Ziyue Wang, Zheao Li, Jinyuan Liu, Yongliang Guan and Chau Yuen are with School of Electrical and Electronics Engineering, Nanyang Technological University, 639798, Singapore (Email:wang1953@e.ntu.edu.sg; zheao001@e.ntu.edu.sg; jinyuan001@e.ntu.edu.sg; EYLGuan@ntu.edu.sg; Chau.Yuen@ntu.edu.sg). \textit{(Corresponding author: Chau Yuen.)}}
\thanks{Xiaoyan Ma is with School of Electrical and Computer Engineering, Purdue University, USA (Email: ma946@purdue.edu). }
\thanks{Xingyu Peng is with College of Information Science and Electronic Engineering, Zhejiang University, Hangzhou 310027, China, and Zhejiang Provincial Key Laboratory of Info. Proc., Commun. \& Netw. (IPCAN), Hangzhou 310027, China (Email: peng\_xingyu@zju.edu.cn).}
}

\markboth{DRAFT}%
{Shell \MakeLowercase{\textit{et al.}}: A Sample Article Using IEEEtran.cls for IEEE Journals}


\maketitle

\begin{abstract}
In this article, we investigate the robust beamforming design for a simultaneous transmitting and reflecting reconfigurable intelligent surface (STAR-RIS) aided downlink rate-splitting multiple access (RSMA) communication system, where both transceivers and STAR-RIS suffer from the impact of hardware impairments (HWI).
A base station (BS) is deployed to transmit messages concurrently to multiple users, utilizing a STAR-RIS to improve communication quality and expand user coverage. We aim to maximize the achievable sum rate of the users while ensuring the constraints of transmit power, STAR-RIS coefficients, and the actual rate of the common stream for all users. To solve this challenging high-coupling and non-convexity problem, we adopt a fractional programming (FP)-based alternating optimization (AO) approach, where each sub-problem is addressed via successive convex approximation (SCA) and penalty function (PF) methods. Numerical results demonstrate that the proposed scheme outperforms other multiple access schemes and conventional passive RIS in terms of the achievable sum rate. Additionally, considering the HWI of the transceiver and STAR-RIS makes our algorithm more robust than when such considerations are not included.
\end{abstract}


\section{Introduction}
\IEEEPARstart{T}{he} rapid expansion of the Internet of Things (IoT) has driven an exponential growth in the number of mobile users and network traffic volumes over the past few decades \cite{A Vision of 6G}, \cite{A Speculative Study on 6G}. Given the limited availability of spectrum resources, next-generation communication networks are urgently seeking new technologies to enhance spectrum efficiency, aiming to achieve higher throughput, enhanced reliability, expanded support for machine-type communications, and broader coverage for both communication and sensing. As a result, researchers are increasingly focused on developing innovative solutions such as network slicing and the non-orthogonal multiple access (NOMA) \cite{Non-orthogonal multiple access},\cite{Network Slicing in 5G}. These advancements are designed to maximize resource utilization, provide seamless connectivity, and enable scalable, flexible networks capable of supporting diverse IoT applications.

Traditional multiple access techniques, such as orthogonal multiple access (OMA), provide feasible solutions for resource distribution in existing multi-user communication systems. However, OMA faces limitations in terms of low spectral efficiency and high-complexity orthogonality restoring measurements\cite{A Survey of Non-Orthogonal Multiple Access for 5G}. Unlike traditional OMA schemes, which allocate resources based on time or frequency, NOMA differentiates users by varying their power levels within the same time and frequency resource block
\citerange{ Nonorthogonal Multiple Access for 5G and Beyond}{ On the Performance of}. In NOMA, each message of the user is encoded into distinct streams, which are superimposed and transmitted simultaneously to all users. At the receiver side, successive interference cancellation (SIC) is used to decode the individual messages based on a predefined decoding order.  Previous studies have demonstrated that NOMA provides improved spectral efficiency and enhanced user fairness in a variety of communication scenarios \cite{ Spectral and Energy-Efficient Wireless},\cite{ Sub-Channel Assignment}. However, NOMA does not fully exploit the spatial dimension. In \cite{ Is NOMA Efficient in Multi-Antenna Networks}, it has been shown that in multiple-input multiple output (MIMO) scenarios, the sum multiplexing gain of space division multiple access (SDMA) is always no less than NOMA. This suggests that NOMA underperforms in terms of sum rate compared to SDMA at high signal-to-noise ratio (SNR) levels, indicating its limited ability to fully exploit the multi-antenna capabilities of communication systems.

Recently, rate-splitting multiple access (RSMA) has emerged as a promising technology for next-generation networks \cite{ Rate-Splitting Multiple Access}. Unlike NOMA and traditional SDMA, the key concept of RSMA is to split the message of each user into a common part and a private part. The common parts of all users are combined and encoded into a common message, while the private parts are separately encoded into individual messages. At the receiver side, users first decode the common message while treating all private streams as interference. Once the common message is successfully decoded, each user then proceeds to decode its own private message, treating the remaining signals from other users as interference. This approach allows interference to be partially treated as noise. By dynamically adjusting the common stream and transmission power, RSMA can adapt to various interference scenarios and further improve spectrum efficiency compared to existing NOMA or SDMA schemes \citerange{ Downlink Energy Efficiency Maximization} { Power Allocation for High}.

Although RSMA has an outstanding and flexible capability for resource allocation, which can enhance spectrum efficiency and resource utilization to a certain extent, the high frequency characteristic in the future network will leads to severe signal fading and poor traversal capability despite the denser and more diverse base station (BS) deployments, which will limiting the full potential of RSMA. To further capitalize on the advantages of RSMA in next-generation networks, the emergence of reconfigurable intelligent surface (RIS) can help to overcome this shortcoming \cite{Towards smart and reconfigurable}. RIS is a reconfigurable planar array composed of numerous passive elements. By controlling the reflection unit array on the RIS surface via a microcontroller, RIS can reflect incoming signals in specific directions, actively shaping the channel environment, enhancing line-of-sight (LoS) components and mitigating signal blind spots. The passive nature of RIS that requires no additional energy beyond the micro-controller, makes it not only cost-effective and easy to deploy, but also low-maintenance. Such characteristics endow RIS with broad application prospects in multiple aspects of communication systems, makes it a highly promising technology for next-generation networks \citerange{Joint Beamforming and Reflecting}{Stacked intelligent metasurfaces-enhanced}. Specifically, in the RSMA transmission scheme, deploying RIS can effectively enhance system performance by providing additional direct transmission paths, which improves both the received signal power at users and the degrees of freedom for beamforming design. Previous research has demonstrated the value of RIS aided RSMA systems. The work \cite{Max-min fair RIS-aided rate-splitting} addressed the maximization of the minimum group rate in multigroup multicast communication scenarios. The work \cite{Robust Weighted Sum-Rate} focused on maximizing the weighted sum rate for multi-user systems, while the work \cite{Robust Beamforming and Rate Optimization} proposed a robust beamforming design for downlink RIS aided symbiotic radio systems with RSMA. The above studies show superior performance for RIS aided RSMA scenarios compared to those without RIS aided.

Traditional RIS can only enable half-space coverage. In fact, current RIS prototypes are limited by hardware and can only serve angles of less than 180 degrees \cite{2-Bit RIS Prototyping Enhancing}. Simultaneous transmitting and reflecting reconfigurable intelligent surfaces (STAR-RIS), a new type of RIS theoretically capable of serving full-space coverage, offers mode switching, energy splitting, and time switching modes \cite{Simultaneously Transmitting and Reflecting}. The hardware implementation of STAR-RIS is different with the conventional passive RIS, as well as its the signal model\cite{Dynamical absorption manipulation} \cite{2-bit intelligent reflection surface enhances}. In the energy-splitting (ES) mode, each element on the STAR-RIS surface has four configurable degrees of freedom to adjust the incident signal: the amplitude and phase of both the reflection and transmission regions. This represents a significant increase in system degrees of freedom and versatility compared to conventional RIS \cite{ A Survey on STAR-RIS: Use Cases}. It surpasses traditional passive RIS in both spatial coverage freedom and the feasible domain of optimization algorithms.

The integrated STAR-RIS and RSMA system, with its 360-degree flexible signal enhancement and excellent interference management, is poised to become a new architectural paradigm for next-generation communication networks. In \cite{Performance Analysis for RSMA}, the authors demonstrated that STAR-RIS empowered RSMA systems achieve lower BS energy consumption at the same outage probability compared to NOMA systems. In \cite{ Sum-Rate Maximization in STAR}, the authors proposed a proximal policy optimization (PPO)-based deep reinforcement learning (DRL) algorithm to solve challenging non-convex optimization problems. The results show that compared to conventional RIS, STAR-RIS significantly enhances system sum rate and user fairness, and the integrated STAR-RIS and RSMA system achieves more notable performance gains than deploying NOMA methods. In \cite{ Secure Transmission Design}, the authors considered the physical layer security performance of downlink integrated STAR-RIS and RSMA systems, using a successive convex approximation (SCA)-based penalization algorithm to achieve significantly enhanced secrecy rates compared to conventional RIS. In \cite{On the Performance Analysis of}, it was demonstrated that STAR-RIS aided RSMA systems also perform excellently in integrated communication and sensing.

Despite previous studies highlight the potential of STAR-RIS aided RSMA systems, perfect hardware implementations at both the transceiver and the STAR-RIS side are typically assumed. In practical communication system applications, hardware degrades over time due to environmental influences, leading to signal distortions that deviate from the intended targets. This degradation affects key components such as power amplifiers, digital-to-analog converters, oscillators, as well as the STAR-RIS itself \cite{Beamforming Optimization for Active RIS-Aided Multiuser}. Specifically, for STAR-RIS, the hardware impairments (HWI) are usually seen as the phase noise among the elements \cite{Achievable Rate Analysis of the STAR-RIS}, which is generated either by phase estimation error or by quantization error. HWI can degrade system performance \citerange{ RIS Assisted Wireless Powered} {Secure Wireless Communication in RIS} and affect the effective multi-user management capabilities of RSMA. Therefore, to ensure robust system performance under realistic deployment conditions, it is necessary to consider the impact of HWI during signal modeling and compensate for their effects during optimization processes using signal processing algorithms. However, considering HWI at the transceiver and STAR-RIS introduces additional variables in the optimization process, complicating signal derivation and increasing algorithmic overhead, making research on HWI consideration more challenging.

In summary, HWI introduces deviations in the signal processing procedure under practical conditions, negatively affecting algorithm performance and significantly degrading system efficiency. Although integrated STAR-RIS and RSMA system offers substantial advantages in various aspects, existing research has not yet fully addressed the impact of HWI in such systems. To bridge this gap, this paper investigates a downlink multi-user scenario that comprehensively incorporates HWI, including impairments at the transceivers and phase noise at the STAR-RIS. This modeling approach better aligns with practical engineering applications and enables optimization and compensation for HWI within the proposed algorithm. The main contributions of this paper are summarized as follows:

\hangafter=1
\setlength{\hangindent}{2em}
•~We investigate a STAR-RIS aided downlink RSMA communication system considering HWI at both the transceivers and the STAR-RIS side. In this setup, the BS transmits both private and common messages to multiple users using RSMA with the help of the STAR-RIS. Then, we aim to maximize the achievable sum rate for the proposed STAR-RIS aided RSMA robust beamforming design by jointly optimizing the transmit beamforming vector at the BS side, the STAR-RIS coefficients, and the actual transmission rate for the common stream. By accounting the HWI, we enhance the STAR-RSMA system to better reflect real-world conditions, compensating for the progressive signal distortion caused by environmental factors over time.

\hangafter=1
\setlength{\hangindent}{2em}
•~To tackle the high coupling and non-convexity optimization problem, we first employ fractional programming (FP) to reformulate the original objective function into a more tractable form. The resulting problem is then decomposed into several sub-problems within an alternating optimization (AO) framework. Furthermore, the transmit beamforming vector and the STAR-RIS coefficients are jointly optimized using a combination of SCA and penalty function (PF) algorithms.

\hangafter=1
\setlength{\hangindent}{2em}
•~Simulation results demonstrate that the proposed FP-AO method achieves rapid convergence and delivers superior performance in the considered scenario. Furthermore, it is verified that the deployment of STAR-RIS significantly outperforms conventional RIS in terms of achievable sum rate performance, yielding up to 23\% higher gain in optimal situations. The robust beamforming design maintains consistent performance advantages over non-robust designs under various HWI conditions. Additionally, RSMA scheme leads to a notable enhancement in communication quality as compared to the NOMA and OMA scheme, while demonstrating stronger resilience against HWIs induced by both transceiver chains and reflecting elements.

\textit{Notation:} In this paper, the notations used are shown as follows. Scalars are denoted by lowercase, vectors in bold lower case, and matrices in bold upper case letters. For the vector $\textbf{x}$, $x_{n}$,$|\textbf{x}|$ and $\text{diag}(\textbf{x})$ represent its $n$-th element, the absolute value of the vector, and the diagonal matrix formed by the vector, respectively. For the matrix $\textbf{X}$, $\textbf{X}^{T}$, $\textbf{X}^{H}$, $\text{Tr}(\textbf{X})$, $|\textbf{X}|_{n,n}$ denotes its transpose, conjugate
transpose, trace, and the element at the $(n,n)$-th position of the matrix. The expectation operator is represented by $\mathbb{E}$[·]. $\odot$ denotes the Hadamard product. $\mathbb{C}^{a\times b}$ represents the set of all complex matrices of size $a$ × $b$. $\widetilde{{\rm{diag}}}$ represents a diagonal matrix whose diagonal entries are the same as those of the input matrix.

\section{System Model And Problem Formulation}
\subsection{System Model}
Fig. 1 shows a STAR-RIS aided downlink RSMA system where the BS is equipped with $M$ antennas, and a STAR-RIS consisting of $N$ elements provides additional signal paths for the users. The STAR-RIS is connected to the BS via a micro-controller, with each element operating in the ES mode. Specifically, each unit on STAR-RIS is capable of simultaneously reflecting and transmitting signals while the power of the output signal is equal to the incident signal, theoretically. Assume there are $K$ users distributed across both the reflection and transmission region of STAR-RIS. Let $\mathcal{T}$ = \{1,\,2,\,.\,.\,.\,,\,$k_{0}$\} and $\mathcal{R}$ = \{$k_{0}+1$,\,$k_{0}+2$,\,.\,.\,.\,,\,$K$\} represent the users in reflection region and transmission region, respectively, where $k_{0}<K$ make sure both region have user located. Then let $\mathcal{K}$ = $\mathcal{T}\cup\mathcal{R}$ be the set of the whole users. By adoping RSMA scheme, for any $k\in \mathcal{K}$, the message sent to the $k$-th user is divided into a common part $m_{k}^{c}$ and a private part $m_{k}^{p}$. The transmitter uses the codebook shared by the BS and all users to merge the common parts $m_{k}^{c}$, into a single stream $s_{c}$ to ensure all the users can decode the message from the common stream. For each $m_{k}^{p}$, a codebook known only to the BS and the $k$-th user is used to encode it into a stream $s_{k}$. Therefore, the BS will continuously transmit a common stream $s_{c}$ and $K$ private streams $s_{k}$ at the same time. To assess the theoretical security capacity of the system, we consider a scenario where complete CSI is available. Given that this assumption can be flexibly relaxed in practical applications, the findings obtained in this paper serve as an upper-bound for practical scenarios. The relevant channel estimation techniques for STAR-RIS aided systems can be
found in \cite{Channel Estimation for STAR-RIS-Aided Wireless Communication}.
Therefore, the transmit signal can be expressed as\begin{equation}
	\textbf{x}=\sum_{k=1}^K \textbf{f}_{k}s_{k}+ \textbf{f}_{c}s_{c}+\textbf{m}_{t},
\end{equation}
\begin{figure}[!t]
	\centering
		\includegraphics[width=3.3in]{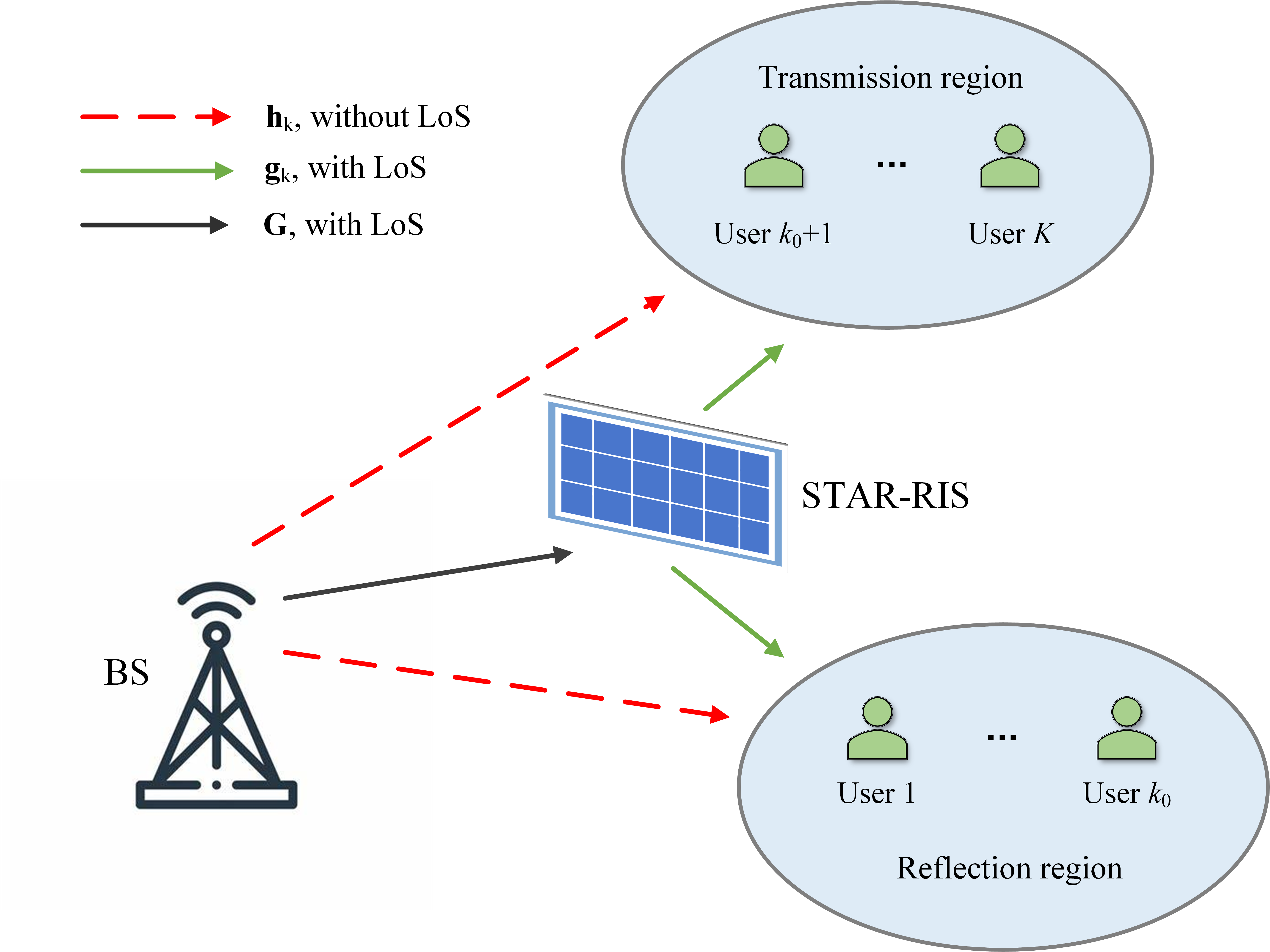}
	\caption{A STAR-RIS aided RSMA communication system with HWI.}
	\label{fig_1}
\end{figure}where $\textbf{f}_{k}\in\mathbb{C}^{M\times1}$, $\textbf{f} _{c}\in\mathbb{C}^{M\times1}$ represents the beamforming vector of private stream and common stream. $\textbf{m}_{t}$ is the HWI variable produced by the transmitter. Previous research have done sufficient investigation both theoretically and practically on this variable considering the joint effect of the component in RF chain and show that it can be modeled as	 $\textbf{m}_{t}\sim\mathcal{C}\mathcal{N}(\textbf{0}, \mu_{t}\widetilde{{\rm{diag}}}(\sum_{k=1}^K\textbf{f}_{k}\textbf{f}_{k}^{H}+\textbf{f}_{c}\textbf{f}_{c}^{H}))$\cite{Secure Wireless Communication in RIS}, where $\mu_{t}\geq0$ is the ratio of transmit distorted noise power to transmit signal power. Let $\mathbb{E}$[$s_{k}s_{k}^{H}$]$=1, \forall k\in\mathcal{K}$.

The channel from BS to the $k$-th user, BS to STAR-RIS and STAR-RIS to the $k$-th user are denoted by $\textbf{h}_{k}\in\mathbb{C}^{M\times1}$, $\textbf{G}\in\mathbb{C}^{N\times M}$ and $\textbf{g}_{k}\in\mathbb{C}^{N\times 1}$, where the $\textbf{h}_{k}$ is without LoS component.\footnote{The algorithm proposed in this work can be extended to general channel conditions, including scenarios where a LoS component exists between the BS and users.} The signal received by the $k$-th user can be represented as
\begin{equation}
\mathit{y}_{k}=\widetilde{\mathit{y}_{k}}+m_{r,k},
\end{equation}
where $m_{r,k}\sim\mathcal{C}\mathcal{N}(0, \mu_{r,k}\mathbb{E}\{|\widetilde{\mathit{y}_{k}}|^{2}\} )$ denotes the HWI at the $k$-th user. $\mu_{r,k}\geq0$ is the ratio of distorted noise power to reveived signal power at the $k$-th user and $\widetilde{\mathit{y}_{k}}$ represents the corresponding received signal: \begin{equation}\widetilde{\mathit{y}_{k}}= \begin{cases}
    (\textbf{h}_{k}^{H}+\textbf{g}_{k}^{H}\mathbf{\Phi }_{t}\mathbf{E}_{t}\textbf{G})\textbf{x}+n_{k},&{k\in\mathcal{T},} \\ 
	(\textbf{h}_{k}^{H}+\textbf{g}_{k}^{H}\mathbf{\Phi }_{r}\mathbf{E}_{r}\textbf{G})\textbf{x}+n_{k},&{k\in\mathcal{R},} 
\end{cases}
\end{equation}
$\mathbf{\Phi}_{i}=\text{diag}(\textbf{v}_{i})$ represents the diagonal reflection matrix of the STAR-RIS, where $\textbf{v}_{i}=[\sqrt{\beta_{1}^{i}}e^{j\theta_{1}^{i}},\sqrt{\beta_{2}^{i}}e^{j\theta_{2}^{i}},\,.\,.\,.\,,\,\sqrt{\beta_{N}^{i}}e^{j\theta_{N}^{i}}]^{T}$ with $i=\{t,r\}$ represents the transmission region and reflection region,  $\sqrt{\beta_{n}^{i}}$ and $\theta _{n}$ being the amplitude and the phase of the $n$-th element, respectively. $\mathbf{E}_{i}=\text{diag}(\bm{\phi}_{i})$ with $\bm{\phi}_{i} = \text{[} e^{j\widetilde{\theta}_{1}^{i}}, e^{j\widetilde{\theta}_{2}^{i}},\,.\,.\,.\,,\,e^{j\widetilde{\theta}_{N}^{i}}  \text{]}^{H}, i=\{t,r\}$ being the phase error vector. It is assumed $\widetilde{\theta}_{n}^{i},n=1 ,2,\,.\,.\,.\,,\,N$ is uniformly distributed in [$-\frac{\pi}{2}, \frac{\pi}{2}$] \cite{RIS Assisted Wireless Powered},\cite{Communication Through a Large Reflecting}. $n_{k}\sim\mathcal{C}\mathcal{N}(0, \sigma_{k}^{2})$ is the additive white gaussian noise (AWGN).

By applying RSMA scheme, we first denote
 \begin{equation}
 	\mathbf{t}_{k}^{H}= \begin{cases}
	\textbf{h}_{k}^{H}+\textbf{g}_{k}^{H}\mathbf{\Phi }_{t}\mathbf{E}_{t}\textbf{G},&{k\in\mathcal{T},}\\ 
	\textbf{h}_{k}^{H}+\textbf{g}_{k}^{H}\mathbf{\Phi }_{r}\mathbf{E}_{r}\textbf{G},&{k\in\mathcal{R},}
\end{cases}
\end{equation}
as the total channel of the $k$-th user. Hence, the achievable rate for the common stream of user $k$ can be expressed as\begin{equation}
	R_{p,k}=\text{log}_{2}(1+ \text{SINR}_{p,k}).
\end{equation}
For private stream, the achievable rate of user $k$ can be expressed as
 \begin{equation}
	R_{c,k}=\text{log}_{2}(1+ \text{SINR}_{c,k}),
\end{equation}
where the $\text{SINR}_{c,k}$ and $\text{SINR}_{p,k}$ are given in (\ref{equation7}) and (\ref{equation8}) at the top of the page.

\begin{figure*}[!t]
	\normalsize
	\setcounter{equation}{6}
	\begin{equation}
		\label{equation7}
	\text{SINR}_{p,k}=\dfrac{ | \textbf{t}_{k}^{H}\textbf{f}_{k}|^{2}}{\sum _{j\neq k}|\textbf{t}_{k}^{H}\textbf{f}_{j}|^{2} + \mu_{r}(\sum _{j=1}^{K}|\textbf{t}_{k}^{H}\textbf{f}_{j}|^{2} + |\textbf{t}_{k}^{H}\textbf{f}_{c}|^{2}) + (1+\mu_{r})\mu_{t}\textbf{t}_{k}^{H}\widetilde{{\rm{diag}}}(\sum_{j=1}^K\textbf{f}_{j}\textbf{f}_{j}^{H}+\textbf{f}_{c}\textbf{f}_{c}^{H})\textbf{t}_{k}+(1+\mu_{r})\sigma_{k}^{2}},
\end{equation}

\begin{equation}
	\label{equation8}
	\text{SINR}_{c,k}=\dfrac{|\textbf{t}_{k}^{H}\textbf{f}_{c}|^{2}}{\sum _{j=1}^{K}|\textbf{t}_{k}^{H}\textbf{f}_{j}|^{2} + \mu_{r}(\sum _{j=1}^{K}|\textbf{t}_{k}^{H}\textbf{f}_{j}|^{2} + |\textbf{t}_{k}^{H}\textbf{f}_{c}|^{2}) + (1+\mu_{r})\mu_{t}\textbf{t}_{k}^{H}\widetilde{{\rm{diag}}}(\sum_{j=1}^K\textbf{f}_{j}\textbf{f}_{j}^{H}+\textbf{f}_{c}\textbf{f}_{c}^{H})\textbf{t}_{k}+(1+\mu_{r})\sigma_{k}^{2}},
\end{equation}
  \rule{\textwidth}{0.4pt}
\end{figure*}

Due to the randomness of the phase matrix generated by the HWI of STAR-RIS, the actual data rate is hard to obtain directly. To tackle this problem, we use the average rate here to approximate $R_{p,k}$ and $R_{c,k}$\cite{Beamforming Optimization for Active RIS-Aided Multiuser},\cite{Achievable rate analysis and phase shift optimization},\cite{Robust transmission design for RIS-assisted secure multiuser} by taking expectation to the phase noise matrix $\mathbf{E}_{i}$ in (7) and (8). The average achievable rate can be represented as
\begin{equation}
	R_{j,k}\approx \widetilde{R}_{j,k} = \text{log}_{2}(1+\widetilde{\gamma}_{j,k}),  j =\{p,c\}.
	\end{equation}
Since $\textbf{t}_{k}$ is the only variable which contains $\mathbf{E}_{i}$, we first address
$\mathbb{E}\{\textbf{t}_{k}\textbf{t}_{k}^{H}\}$ as follows:
\begin{flalign}
			&\mathbb{E}\{\textbf{t}_{k}\textbf{t}_{k}^{H}\}=\mathbf{G}^{H}\mathbf{\Phi}^{H}_{t}\text{diag}(\mathbf{g}_{k})\mathbb{E}\{\bm{\phi}_{t}\bm{\phi}^{H}_{t}\}\text{diag}(\mathbf{g}_{k}^{H})\mathbf{\Phi}_{t} \mathbf{G} \notag\\&
	+\mathbf{G}^{H}\mathbf{\Phi}^{H}_{t}\text{diag}(\mathbf{g}_{k})\mathbb{E}\{\bm{\phi}^{\ast}_{t}\}\textbf{h}_{k}^{H}
	+\textbf{h}_{k}\mathbb{E}\{\bm{\phi}^{\ast}_{t}\}\text{diag}(\mathbf{g}_{k}^{H})\mathbf{\Phi}_{t}\textbf{G}
	\notag\\&+ 
	\textbf{h}_{k}\textbf{h}_{k}^{H}, {k\in\mathcal{T},} \label{10}
\end{flalign}
\begin{flalign}
	&\mathbb{E}\{\textbf{t}_{k}\textbf{t}_{k}^{H}\}=\mathbf{G}^{H}\mathbf{\Phi}^{H}_{r}\text{diag}(\mathbf{g}_{k})\mathbb{E}\{\bm{\phi}_{r}\bm{\phi}^{H}_{r}\}\text{diag}(\mathbf{g}_{k}^{H})\mathbf{\Phi}_{r} \mathbf{G} \notag\\&
	+\mathbf{G}^{H}\mathbf{\Phi}^{H}_{r}\text{diag}(\mathbf{g}_{k})\mathbb{E}\{\bm{\phi}^{\ast}_{r}\}\textbf{h}_{k}^{H}
	+\textbf{h}_{k}\mathbb{E}\{\bm{\phi}^{\ast}_{r}\}\text{diag}(\mathbf{g}_{k}^{H})\mathbf{\Phi}_{r}\textbf{G}
	\notag\\&+ 
	\textbf{h}_{k}\textbf{h}_{k}^{H}, {k\in\mathcal{R}.} \label{11}
\end{flalign}

In order to calculate
$\mathbb{E}\{\bm{\phi}_{i}\bm{\phi}^{H}_{i}\}$ and  $\mathbb{E}\{\bm{\phi}^{\ast}_{i}\}$, we denote $\delta_{\theta}=\widetilde{\theta} _{i} -\widetilde{\theta} _{j}$,$\forall i,j =1,\,2,\,.\,.\,.\,,\,N$. The probability function of $\widetilde{\theta} _{n}$ can be express as $f(\widetilde{\theta} _{n})=\frac{1}{\pi}$. Thus,
\begin{equation}
\delta_{\theta} = \begin{cases}
	\dfrac{1}{\pi ^{2}}\delta_{\theta}+ \dfrac{1}{\pi},&\delta_{\theta}\in[-\pi,0], \\
	-\dfrac{1}{\pi ^{2}}\delta_{\theta}+ \dfrac{1}{\pi},&\delta_{\theta}\in[0,\pi],
		\end{cases}
\end{equation}
and since we have $\mathbb{E}_{\delta_{\theta}}\{e^{j\delta_{\theta}}\} = \mathbb{E}_{\delta_{\theta}}\{e^{j\theta_{i}-j\theta_{j}}\} = \int_{-\pi}^{\pi}f(\delta_{\theta})e^{j\delta_{\theta}}d\delta_{\theta} =\frac{4}{\pi^{2}}$, we can give the expression 
\begin{equation}
 \mathbb{E}\{\bm{\phi}_{i}\bm{\phi}^{H}_{i}\} = \textbf{I}_{N}+ \textbf{J} = \left ({{ \begin{matrix} 1& \cdots & \frac{4}{\pi^{2}} \\ \frac{4}{\pi^{2}}& \cdots & \frac{4}{\pi^{2}} \\ \vdots & \ddots & \vdots \\ \frac{4}{\pi^{2}}& \cdots & 1 \\ \end{matrix} }}\right ), \label{13}
\end{equation}
where \textbf{J} is a $N\times N$ matrix
\begin{equation}
\textbf{J}_{(i_{1},i_{2})} =\begin{cases}
	0, &i_{1}=i_{2},  \\
	\frac{4}{\pi^{2}}, &i_{1}\neq i_{2}.
	\end{cases}
\end{equation}
For $\mathbb{E}\{\bm{\phi}^{\ast}_{i}\}$, we have $\mathbb{E}_{\theta_{i}}\{e^{-j\theta_{i}}\} = \int_{-\frac{\pi}{2}}^{\frac{\pi}{2}}f(\theta_{i})(\cos\theta_{i} - j\sin\theta_{i})d\theta_{i} = \frac{2}{\pi}$. Then, we can have
\begin{equation}
\mathbb{E}\{\bm{\phi}^{\ast}_{i}\}= \dfrac{2}{\pi} \textbf{1}_{N}, \label{15}
\end{equation}
where $\textbf{1}_{N} $ is a column vector with all elements being one. By applying (\ref{13}) and (\ref{15}), (\ref{10}) and (\ref{11}) can be replaced by 
\begin{equation}
\begin{split}
&\mathbb{E}\{\textbf{t}_{k}\textbf{t}_{k}^{H}\} =\begin{cases} 
	&\mathbf{G}^{H}\mathbf{\Phi}_{t}^{H}\text{diag}(\mathbf{g}_{k})\text{\textbf{DD}}^{T}\text{diag}(\mathbf{g}_{k}^{H})\mathbf{\Phi}_{t} \mathbf{G} +\\&(\frac{2}{\pi}\mathbf{G}^{H}\mathbf{\Phi}_{t}^{H}\mathbf{g}_{k}+\mathbf{h}_{k}) 
	(\frac{2}{\pi}\mathbf{g}_{k}^{H}\mathbf{\Phi}_{t}\mathbf{G}+\mathbf{h}_{k}^{H}),{k\in\mathcal{T}}\\ &
	\mathbf{G}^{H}\mathbf{\Phi}_{r}^{H}\text{diag}(\mathbf{g}_{k})\text{\textbf{DD}}^{T}\text{diag}(\mathbf{g}_{k}^{H})\mathbf{\Phi}_{r} \mathbf{G} +\\&(\frac{2}{\pi}\mathbf{G}^{H}\mathbf{\Phi}_{r}^{H}\mathbf{g}_{k}+\mathbf{h}_{k}) 
	(\frac{2}{\pi}\mathbf{g}_{k}^{H}\mathbf{\Phi}_{r}\mathbf{G}+\mathbf{h}_{k}^{H}),{k\in\mathcal{R}}\\ 
\end{cases}\\& = \overline {\mathbf {T}} _{k},
\end{split}
\end{equation}
where $\textbf{D}\textbf{D}^{T}\triangleq (1-\frac{4}{\pi^{2}}) \textbf{I}_{N}$. Now, the exact expressions of $\widetilde{\gamma}_{p,k}$ and $\widetilde{\gamma}_{c,k}$ can be expressed as
\begin{equation}
	 \widetilde{\gamma}_{p,k} = \dfrac{\text{Tr}\big(\overline {\mathbf {T}} _{k}\textbf{f}_{k}\textbf{f}_{k}^{H}\big)}{\text{Tr}\big(\overline {\mathbf {T}} _{k}(\textbf{F}_{a}+\textbf{F}_{b})\big)-\text{Tr}\big(\overline {\mathbf {T}} _{k}\textbf{f}_{k}\textbf{f}_{k}^{H}\big)+(1+\mu_{r})\sigma^{2}_{k}},
\end{equation}
\begin{equation}
	\widetilde{\gamma}_{c,k} = \dfrac{\text{Tr}\big(\overline {\mathbf {T}} _{k}\textbf{f}_{c}\textbf{f}_{c}^{H}\big)}{\text{Tr}\big(\overline {\mathbf {T}} _{k}(\textbf{F}_{a}+\textbf{F}_{b})\big)+(1+\mu_{r})\sigma^{2}_{k}},
\end{equation}
where $\textbf{F}_{a} = \mu_{r}\textbf{f}_{c}\textbf{f}_{c}^{H}+(1+\mu_{r})\mu_{t}\widetilde{\text{diag}}(\textbf{f}_{c}\textbf{f}_{c}^{H})$ and $\textbf{F}_{b} = (1+\mu_{r})\big(\textbf{F}\textbf{F}^{H}+\mu_{t}\widetilde{\text{diag}}(\textbf{F}\textbf{F}^{H})\big)$, $\textbf{F} = \text{[} \textbf{f}_{1}, \textbf{f}_{2},\,.\,.\,.\,,\,\textbf{f}_{k} \text{]} \in\mathbb{C}^{M\times K} $. Furthermore, the average achievable private rate (AAPR) $R_{p,k}$ and the average achievable common rate (AACR) $R_{c,k}$ can be expressed as
\begin{equation}
\widetilde{R}_{j,k} = \text{log}_{2}(1+\widetilde{\gamma}_{j,k}),  j =\{p,c\}.
\end{equation}

Here, to make sure that all the users can successfully decode the common part of the message, the actual transmission rate for the common stream $\textbf{c} = \text{[}c_{1}, c_{2},\,.\,.\,.\,,\,c_{K} \text{]}$ should satisfy that $	\sum_{k=1}^{K}c_{k}\leq \underset{k\in\mathcal{K}}{\text{min}} R_{c,k}$, where $c_{k}$ is the actual common rate allocated for the $k$-th user.
\subsection{Problem Formulation}
Considering the STAR-RIS aided downlink RSMA communication system with HWI, our objective is to maximize the overall achievable sum rate by jointly optimizing the beamforming vectors at the base station and STAR-RIS, as well as the actual transmission rate allocated to the common stream for each users. The optimization problem is formulated as follows:
\begin{subequations}
\begin{flalign}
(P_{0}):& \underset{\mathbf{f}_{k},\mathbf{f}_{c},\mathbf{\Phi} _{i}, \textbf{c}}{\text{max}} \sum _{k=1}^{K}(\widetilde{R}_{p,k} + c_{k}), \label{objective function 0}\\
\text{s.t.} \ \ &\sum _{k=1}^{K}||\mathbf{f}_{k}||^{2}+||\mathbf{f}_{c}||^{2}\leq P_{max} \label{Pmax},\\
&|\mathbf{\Phi} _{t}|_{n,n}^{2} + |\mathbf{\Phi }_{r}|_{n,n}^{2} = 1, \forall n, \label{RIS phase}\\
&\sum_{k=1}^{K}c_{k}\leq \widetilde{R}_{c,k} ,\forall k,  \label{c1}\\
&c_{k}\geq 0, \forall k.  \label{c2}
\end{flalign}
\end{subequations}
Constraint (\ref{Pmax}) ensures that the transmit power remains within the allowable range of the maximum power budget $P_{max}$. Constraint (\ref{RIS phase}) indicates the amplitude of  STAR-RIS elements should meet the law of energy conservation and the phase should satisfy the rules of the ES mode. Constraint (\ref{c1}) guarantee the successfully decoding of the common stream, and (\ref{c2}) illustrates that the actual achievable common rate should be non-negative.

\section{Proposed algorithm }
In this section we propose a FP-AO algorithm to solve $(P_{0})$. To deal with the highly complicated fractional term objective function, we first transform the original problem into a more tractable form based on the theory of FP. Then we alternatively optimize the auxiliary variables introduced by FP, the BS beamforming vector for both common and private stream as well as the actual common rate allocated for users, and the phase of the STAR-RIS.
\subsection{Problem Reformulation by FP}
The objective function of $(P_{0})$ is hard to tackle due to the sum of multiple logarithmic functions. Motivated by previous works\cite{Fractional programming for communication systems-Part I}, \cite{Fractional programming for communication systems-Part II}, \cite{Reflection and Relay Dual-Functional RIS}, the closed-form FP techniques is used to simplify this sum-of-log-of-ratio problem. We first utilize lagrangian dual transform to deal with the log function by introducing auxiliary variable $\textbf{a} = \text{[}a_{1}, a_{2},\,.\,.\,.\,,\,a_{K} \text{]}$, thereby $(P_{0})$ can be reformulated as

\begin{flalign}
	(P_{1}): \  \underset{\mathbf{f}_{k},\mathbf{f}_{c},\mathbf{\Phi} _{i}, \textbf{a},\textbf{c}}{\text{max}} g_{R}(\mathbf{f}_{k},\mathbf{f}_{c},\mathbf{\Phi }_{i}, \textbf{a},\textbf{c}), \label{objective function P1} \\
	\text{s.t.} \ \ \ \ \text{(\ref{Pmax})},\text{(\ref{RIS phase})},\text{(\ref{c1})},\text{(\ref{c2}),}  \ \ \ \notag
\end{flalign}
where the objective function
\begin{flalign} g_{R}(&\mathbf{f}_{k},\mathbf{f}_{c},\mathbf{\Phi }_{i}, \textbf{a},\textbf{c}) = \sum_{k=1}^{K}\Big( \text{ln}(1+a_{k}) - a_{k} \notag  \\ &+\dfrac{(1+a_{k}) \text{Tr}\big(\overline {\mathbf {T}} _{k}\textbf{f}_{k}\textbf{f}_{k}^{H}\big)}{\text{Tr}\big(\overline {\mathbf {T}} _{k}(\textbf{F}_{a}+\textbf{F}_{b})\big)+(1+\mu_{r})\sigma^{2}_{k}}\Big)+\sum_{k=1}^{K} c_{k}. \label{auxiliary1}
\end{flalign}
	However, the problem $(P_{1})$ is still hard to tackle since the presence of the sum of fractional terms. Upon applying quadratic transform  \cite{Fractional programming for communication systems-Part II} by introducing the auxiliary variable $\textbf{b} = \text{[}b_{1}, b_{2},\,.\,.\,.\,,\,b_{K} \text{]}$, the problem $(P_{1})$ can be further transformed into
	\begin{figure*}[!t]
		\normalsize
		\setcounter{equation}{22}
		\begin{equation}
			\label{objective function after FP}
			f_{R}(\mathbf{f}_{k},\mathbf{f}_{c},\mathbf{\Phi }_{i}, \textbf{a},\textbf{b},\textbf{c})=\sum_{k=1}^{K}\bigg(\text{ln}\big(1+a_{k}\big)-a_{k} +2b_{k}\sqrt{(1+a_{k})\text{Tr}\big(\overline {\mathbf {T}} _{k}\textbf{f}_{k}\textbf{f}_{k}^{H}\big)}  -b_{k}^{2}\Big(\text{Tr}\big(\overline {\mathbf {T}} _{k}(\textbf{F}_{a}+\textbf{F}_{b})\big)+(1+\mu_{r})\sigma^{2}_{k}\Big)+c_{k}\bigg) , 
		\end{equation} 
		
		\rule{\textwidth}{0.4pt} 
	\end{figure*}
\begin{flalign}
		(\widehat{P_{1}}): \  \underset{\mathbf{f}_{k},\mathbf{f}_{c},\mathbf{\Phi }_{i}, \textbf{a},\textbf{b},\textbf{c}}{\text{max}} f_{R}(\mathbf{f}_{k},\mathbf{f}_{c},\mathbf{\Phi }_{i}, \textbf{a},\textbf{b},\textbf{c}), \label{objective function P2} \\
		\text{s.t.} \ \ \ \ \ \ \ \ \  \text{(\ref{Pmax})},\text{(\ref{RIS phase})},\text{(\ref{c1})},\text{(\ref{c2}),	}  \  \ \ \ \ \ \notag
\end{flalign}
where $f_{R}(\mathbf{f_{k}},\mathbf{f_{c}},\mathbf{\Phi _{i}}, \textbf{a},\textbf{b},\textbf{c})$ is formulated as (\ref{objective function after FP}) at the top of the next page. 

Compared with the original problem, the transferred form is more tractable by introducing two auxiliary variables. To solve $(\widehat{P_{1}})$, we capitalize SCA and PF method alternatively to each group of variables while others kept fixed. The sub-problems of updating each group are specified in details in the following subsections.

\subsection{Update the Auxiliary Variable Vectors}
When variables $\mathbf{f}_{k},\mathbf{f}_{c},\mathbf{\Phi }_{i}, \textbf{c}$ are fixed, the auxiliary variable vectors $\textbf{a},\textbf{b}$ introduced by FP will be easy to optimize by simply take partial derivative of (\ref{auxiliary1}) and (\ref{objective function after FP}) and follow the equation below: 
\begin{equation}
\frac{\partial g_{R}(\mathbf{f}_{k},\mathbf{f}_{c},\mathbf{\Phi }_{i}, \textbf{a},\textbf{c})}{\partial a_{k}} = 0,
\end{equation}

\begin{equation}
\frac{\partial	f_{R}(\mathbf{f}_{k},\mathbf{f}_{c},\mathbf{\Phi }_{i}, \textbf{a},\textbf{b},\textbf{c})}{\partial b_{k}} = 0.
\end{equation}
Therefore, the closed-form of the optimal $a_{k}^{\ast},b_{k}^{\ast}$ are given as

\begin{equation}
a_{k}^{\ast} = \widetilde{\gamma}_{p,k},
\end{equation}

\begin{equation}
b_{k}^{\ast} = 	\dfrac{\sqrt{(1+a_{k})\text{Tr}\big(\overline {\mathbf {T}} _{k}\textbf{f}_{k}\textbf{f}_{k}^{H}\big)}}{\text{Tr}\big(\overline {\mathbf {T}} _{k}(\textbf{F}_{a}+\textbf{F}_{b})\big)+(1+\mu_{r})\sigma_{k}^{2}}.
\end{equation}

\subsection{Update the BS Beamforming and the Actual Common Rate}
For fixed $\textbf{a},\textbf{b}$ and $\mathbf{\Phi}_{i}$ , we focus on the optimization of $\{\mathbf{f}_{k},\mathbf{f}_{c}, \textbf{c}\} $. By omitting the constant value in the objective function, the corresponding optimization problem can be rewritten as
\begin{flalign}
		(P_{2}): \ \ \ & \underset{\mathbf{f}_{k},\mathbf{f}_{c},\textbf{c}}{\text{max}} \ \  \sum_{k=1}^{K}\bigg(2b_{k}\sqrt{(1+a_{k})\text{Tr}\big(\overline {\mathbf {T}} _{k}\mathbf{f}_{k}\mathbf{f}_{k}^{H}\big)} \notag \\  & \ \ \ \ \ \ \ \ -b_{k}^{2}\text{Tr}\big(\overline {\mathbf {T}} _{k}(\textbf{F}_{a}+\textbf{F}_{b})\big)+c_{k}\bigg), \label{objective function P2.1} \\ &
		\text{s.t.} \ \ \  \ \  \text{(\ref{Pmax})},\text{(\ref{c1})},\text{(\ref{c2}).	} \notag
 \end{flalign}
To further simplify the problem, let define  $\widetilde{\textbf{F}}_{k}=\textbf{f}_{k}\textbf{f}_{k}^{H}$ and $\widetilde{\textbf{F}}_{c}=\textbf{f}_{c}\textbf{f}_{c}^{H}$, which should satisfy the constraints $\widetilde{\textbf{F}}_{k} \succeq \textbf{0}$, $\widetilde{\textbf{F}}_{c} \succeq \textbf{0}$ and $\text{rank} (\widetilde{\textbf{F}}_{k})= \text{rank} (\widetilde{\textbf{F}}_{c})=\text{1}$. However, the rank-1 constraint is naturally satisfied here \cite{Energy efficiency maximization via joint}. Hence, $\textbf{F}_{a}$ and $\textbf{F}_{b}$ can be rewritten as
\begin{equation}
\textbf{F}_{a}' = \mu_{r}\widetilde{\textbf{F}}_{c}+(1+\mu_{r})\mu_{t}\widetilde{\text{diag}}(\widetilde{\textbf{F}}_{c}),
 \end{equation}
\begin{equation}
\textbf{F}_{b}' = (1+\mu_{r})\big(\sum_{k=1}^{K}\widetilde{\textbf{F}}_{k}+\mu_{t}\widetilde{\text{diag}}(\sum^{K}_{k=1}\widetilde{\textbf{F}}_{k})\big).
\end{equation}
As a result, the objective function (\ref{objective function P2.1})  can be rewritten as

\begin{flalign}
f_{R2}(\widetilde{\mathbf{F}}_{k},&\widetilde{\mathbf{F}}_{c},\textbf{c})=\sum_{k=1}^{K}\bigg(2b_{k}\sqrt{(1+a_{k})\text{Tr}\big(\overline {\mathbf {T}} _{k}\widetilde{\textbf{F}}_{k}\big)} \notag \\-&b_{k}^{2}\text{Tr}\big(\overline {\mathbf {T}} _{k}(\textbf{F}_{a}'+\textbf{F}_{b}')\big)+c_{k}\bigg),
\end{flalign}
and the AACR can be expressed as
\begin{flalign}
\widetilde{R}_{c,k} &= \text{log}_{2}\Big(\gamma_{k}\text{Tr}\big(\overline {\mathbf {T}} _{k}(\textbf{F}_{a}'+\textbf{F}_{b}')\big)+\gamma_{k}\text{Tr}\big(\overline {\mathbf {T}} _{k}\widetilde{\textbf{F}}_{c}\big)+1\Big)\notag \\&- \text{log}_{2}\Big(\gamma_{k}\text{Tr}\big(\overline {\mathbf {T}} _{k}(\textbf{F}_{a}'+\textbf{F}_{b}')\big)+1\Big),
\end{flalign}
where we let $\gamma_{k}=\frac{1}{(1+\mu_{r})\sigma^{2}_{k}}$ for convenience. Hence, the problem is reformulated as 
\begin{subequations}
\begin{flalign}	(\overline{P}_{2}): \underset{\widetilde{\mathbf{F}}_{k},\widetilde{\mathbf{F}}_{c},\textbf{c}}{\text{max}}& f_{R2}(\widetilde{\mathbf{F}}_{k},\widetilde{\mathbf{F}}_{c},\textbf{c}), \label{objective function P2.2} \\
	\text{s.t.}\quad \
	 &\text{(\ref{c2})},\notag\\
	 &\widetilde{\textbf{F}}_{k} \succeq \textbf{0},\widetilde{\textbf{F}}_{c} \succeq \textbf{0},\label{sedi}\\
	 &\text{Tr}(\sum^{K}_{k=1}\widetilde{\textbf{F}}_{k}+\widetilde{\textbf{F}}_{c})\leq P_{max}, \label{P2_constraint2} \\
	&\sum_{k=1}^{K}c_{k}\leq \widetilde{R}_{c,k} ,\forall k. \label{c1.2} 
\end{flalign}
\end{subequations}
However, the problem $(\overline{P}_{2})$ is still hard to solve since the non-convex constraint (\ref{c1.2}). 
According to \cite{Secure Wireless Communication in RIS}, (\ref{c1.2}) can be transformed into a tractable form by leveraging the SCA method. Specifically, by adopting the first-order Taylor approximation, the right-hand-side of (\ref{c1.2}) can be approximated as
\begin{flalign}
	\widetilde{R}_{c,k} &\geq \text{log}_{2}\Big(\gamma_{k}\text{Tr}\big(\overline {\mathbf {T}} _{k}(\textbf{F}_{a}'+\textbf{F}_{b}')\big)+\gamma_{k}\text{Tr}\big(\overline {\mathbf {T}} _{k}\widetilde{\textbf{F}}_{c}\big)+1\Big)\notag \\&- \text{log}_{2}\Big(\gamma_{k}\text{Tr}\big(\overline {\mathbf {T}} _{k}(\textbf{F}_{a}^{n}+\textbf{F}_{b}^{n})\big)+1\Big)\notag\\&-
	\dfrac{\gamma_{k}\text{Tr}\big( \overline{\textbf{T}}_{k}(\textbf{F}_{a}'-\textbf{F}_{a}^{n}+\textbf{F}_{b}'-\textbf{F}_{b}^{n})\big)}{\text{ln2}\big(\gamma_{k}\text{Tr}(\overline{\textbf{T}}_{k}(\textbf{F}_{a}^{n}+\textbf{F}_{b}^{n}))  +1\big)} \triangleq \overline{R}_{c,k},
\end{flalign}
\begin{algorithm}
	\caption{FP-AO Algorithm}\label{alg:cap}
	\begin{algorithmic}[1]
		\Require $\mathbf{f}_{k},\mathbf{f}_{c},\mathbf{\Phi} _{i}, \textbf{c}$
		\State Use FP to transfer (\ref{objective function 0}) to (\ref{objective function after FP}) by introduce $\textbf{a}$ and $\textbf{b}$.
		\Repeat 
		\State Update $\textbf{a}, \textbf{b}$.
		\State Update $\textbf{f}_{k}, \textbf{f}_{c}, \textbf{c}$ by solving problem $\widehat{P_{2}}$.
		\State Update $\mathbf{\Phi}_{i}$ by solving problem $\widehat{P_{3}}$. 
		\Until The objective function (\ref{objective function 0}) convergence.
	\end{algorithmic}
\end{algorithm}
where $\textbf{F}_{a}^{n}$ and $\textbf{F}_{b}^{n}$ are the solutions obtained in the $n$-th iteration. Hence, the problem $(\overline{P}_{2})$ can be transformed into
\begin{subequations}
	\begin{flalign}
		(\widehat{P_{2}}): \underset{\widetilde{\mathbf{F}}_{k},\widetilde{\mathbf{F}}_{c},\textbf{c}}{\text{max}}& f_{R2}(\widetilde{\mathbf{F}}_{k},\widetilde{\mathbf{F}}_{c},\textbf{c}),  \\
		\text{s.t.}\quad
		&\text{(\ref{c2})},\text{(\ref{sedi})},\text{(\ref{P2_constraint2})}\notag\\
		&\sum_{k=1}^{K}c_{k}\leq \overline{R}_{c,k} ,\forall k. \label{c1.2.2} 
	\end{flalign}
\end{subequations}
At this stage, all the constraints are convex and the problem can be solved by CVX tool.

\subsection{Update the Phase Matrix of STAR-RIS}
For fixed $\textbf{a}, \textbf{b}, \textbf{f}_{c}, \textbf{f}_{k}$ and $ \textbf{c}$, we aim to optimize the phase shift matrix $\mathbf{\Phi} _{i}$ of the STAR-RIS. By omitting the constant value in the objective function, the corresponding optimization problem can be expressed as
\begin{flalign}
	(P_{3}): \ \ \ & \underset{\mathbf{\Phi}_{i}}{\text{max}} \ \  \sum_{k=1}^{K}\bigg(2b_{k}\sqrt{(1+a_{k})\text{Tr}\big(\overline {\mathbf {T}} _{k}\mathbf{f}_{k}\mathbf{f}_{k}^{H}\big)} \notag \\  & \ \ \ \ \ \ \ \ -b_{k}^{2}\text{Tr}\big(\overline {\mathbf {T}} _{k}(\textbf{F}_{a}+\textbf{F}_{b})\big)\bigg), \label{objective function P3.1} \\ & \ 
	\text{s.t.} \ \ \  \ \  \text{(\ref{RIS phase})},\text{(\ref{c1})}. \notag
\end{flalign}
By observing (\ref{objective function P3.1}), we first extract and handle $\text{Tr}\big(\overline {\mathbf {T}} _{k}\textbf{f}_{k}\textbf{f}_{k}^{H}\big)$ and $\text{Tr}\big(\overline {\mathbf {T}} _{k}(\textbf{F}_{a}+\textbf{F}_{b})\big)$, as  these are the only two parts containing the target variables to be optimized.

For $\overline {\mathbf {T}}_{k}$, let define $\overline {\mathbf {T}}_{k,t}$ and $\overline {\mathbf {T}}_{k,r}$ as the cases for $k\in\mathcal{T}$ and $k\in\mathcal{R}$, respectively. Considering the same form of these two variables, we first address $\overline {\mathbf {T}}_{k,t}$ by rewriting it as follows
\begin{flalign}
	\text{Tr}\big(\overline {\mathbf {T}} _{k,t}\textbf{f}_{k}\textbf{f}_{k}^{H}\big)&= \text{Tr}\big(\mathbf{G}^{H}\mathbf{\Phi}_{t}^{H}\text{diag}(\mathbf{g}_{k})\textbf{V}_{x}\text{diag}(\mathbf{g}_{k}^{H})\mathbf{\Phi}_{t} \mathbf{G}\textbf{f}_{k}\textbf{f}_{k}^{H}\big) \notag \\&+
	\frac{2}{\pi}\textbf{f}_{k}^{H}\textbf{G}^{H}\mathbf{\Phi}_{t}^{H}\textbf{g}_{k}\textbf{h}_{k}^{H}\textbf{f}_{k} +\frac{2}{\pi}\textbf{f}_{k}^{H}\textbf{h}_{k}\textbf{g}_{k}^{H}\mathbf{\Phi}_{t}\textbf{G}\textbf{f}_{k}\notag \\& +\textbf{f}_{k}^{H}\textbf{h}_{k}\textbf{h}_{k}^{H}\textbf{f}_{k}, \label{phase matrix term 1}
	\end{flalign}
where 
\begin{flalign}  \textbf{V}_{x}=\textbf{I}_{N}+ \textbf{J}. \end{flalign}
Then, each term in (\ref{phase matrix term 1}) is handled individually. For the first term we have 

\begin{flalign}
\text{Tr}\big(\mathbf{G}^{H}\mathbf{\Phi}_{t}^{H}\text{diag}(\mathbf{g}_{k})\textbf{V}_{x}\text{diag}(\mathbf{g}_{k}^{H})\mathbf{\Phi}_{t} \mathbf{G}\textbf{f}_{k}\textbf{f}_{k}^{H}\big)=\textbf{v}_{t}^{H}\textbf{Q}_{k}\textbf{v}_{t},
	\end{flalign}
where $\textbf{Q}_{k}=\big(\text{diag}(\textbf{g}_{k})\textbf{V}_{x}\text{diag}(\textbf{g}_{k}^{H})\big) \odot\big(\textbf{G}\textbf{f}_{k}\textbf{f}_{k}^{H}\textbf{G}^{H}\big)^{T}$. For the second term we have
\begin{flalign}	\frac{2}{\pi}\textbf{f}_{k}^{H}\textbf{G}^{H}\mathbf{\Phi}_{t}^{H}\textbf{g}_{k}\textbf{h}_{k}^{H}\textbf{f}_{k}=\textbf{v}_{t}^{H}\textbf{q}_{k},
	\end{flalign}
where $\textbf{q}_{k} =\frac{2}{\pi}\text{diag}(\textbf{h}_{k}^{H}\textbf{f}_{k}\textbf{f}_{k}^{H}\textbf{G}^{H})\textbf{g}_{k}$. The third term is equal to the conjugate transpose of the second which can be represented as
\begin{flalign}
	\frac{2}{\pi}\textbf{f}_{k}^{H}\textbf{h}_{k}\textbf{g}_{k}^{H}\mathbf{\Phi}_{t}\textbf{G}\textbf{f}_{k}=\textbf{q}_{k}^{H}\textbf{v}_{t}.
	\end{flalign}
From above, the transformed term of (\ref{phase matrix term 1}) can be expressed as
\begin{flalign}
\text{Tr}\big(\overline {\mathbf {T}} _{k,t}\textbf{f}_{k}\textbf{f}_{k}^{H}\big)&=\textbf{v}_{t}^{H}\textbf{Q}_{k}\textbf{v}_{t}+\textbf{v}_{t}^{H}\textbf{q}_{k}+\textbf{q}_{k}^{H}\textbf{v}_{t}\notag\\&+\textbf{f}_{k}^{H}\textbf{h}_{k}\textbf{h}_{k}^{H}\textbf{f}_{k}. \label{phase matrix term 2}
		\end{flalign}		
To further transform the expression, we define two extended matrix
\begin{flalign}
 \overline{\textbf{v}}_{t}=\begin{bmatrix}
 	\textbf{v}_{t} \\
 	1
 \end{bmatrix}
 =
 \begin{bmatrix}
 \textbf{v}_{t}^{H}&1
 \end{bmatrix}^{H},
\end{flalign}
and
\begin{flalign}
	\overline{\textbf{Q}}_{k}=\begin{bmatrix}
		\textbf{Q}_{k} & \textbf{q}_{k} \\
		\textbf{q}_{k}^{H} & \textbf{f}_{k}^{H}\textbf{h}_{k}\textbf{h}_{k}^{H}\textbf{f}_{k}
	\end{bmatrix}.
\end{flalign}	
Therefore, 
\begin{flalign}
	\text{Tr}\big(\overline {\mathbf {T}} _{k,t}\textbf{f}_{k}\textbf{f}_{k}^{H}\big) =\overline{\textbf{v}}_{t}^{H}\overline{\textbf{Q}}_{k}\overline{\textbf{v}}_{t}=\text{Tr}(\overline{\textbf{Q}}_{k}\overline{\textbf{V}}_{t}),
\end{flalign}	
where $\overline{\textbf{V}}_{t}=\overline{\textbf{v}}_{t}\overline{\textbf{v}}_{t}^{H}$.

Since $\text{Tr}\big(\overline {\mathbf {T}} _{k,t}\textbf{f}_{k}\textbf{f}_{k}^{H}\big)$, $\text{Tr}\big(\overline {\mathbf {T}} _{k,r}\textbf{f}_{k}\textbf{f}_{k}^{H}\big)$,  $\text{Tr}\big(\overline {\mathbf {T}} _{k,t}(\textbf{F}_{a}+\textbf{F}_{b})\big)$ and $\text{Tr}\big(\overline {\mathbf {T}} _{k,r}(\textbf{F}_{a}+\textbf{F}_{b})\big)$ have similar forms, we can easily derive
\begin{flalign}
		\text{Tr}\big(\overline {\mathbf {T}} _{k,r}\textbf{f}_{k}\textbf{f}_{k}^{H}\big) =\text{Tr}(\overline{\textbf{Q}}_{k}\overline{\textbf{V}}_{r}),
\end{flalign}	
where $\overline{\textbf{V}}_{r}=\overline{\textbf{v}}_{r}\overline{\textbf{v}}_{r}^{H}$, and
\begin{flalign}
	\text{Tr}\big(\overline {\mathbf {T}} _{k,t}(\textbf{F}_{a}+\textbf{F}_{b})\big) = \text{Tr}(\overline{\textbf{S}}_{k}\overline{\textbf{V}}_{t}),
\end{flalign}	
\begin{flalign}
	\text{Tr}\big(\overline {\mathbf {T}} _{k,r}(\textbf{F}_{a}+\textbf{F}_{b})\big) = \text{Tr}(\overline{\textbf{S}}_{k}\overline{\textbf{V}}_{r}),
\end{flalign}	
where
\begin{flalign}
		\overline{\textbf{S}}_{k}=\begin{bmatrix}
		\textbf{S}_{k} & \textbf{s}_{k} \\
		\textbf{s}_{k}^{H} & \textbf{h}_{k}^{H}(\textbf{F}_{a}+\textbf{F}_{b})\textbf{h}_{k}
	\end{bmatrix},
\end{flalign}	
with 
\begin{flalign}
\textbf{S}_{k}=\big(\text{diag}(\textbf{g}_{k})\textbf{V}_{x}\text{diag}(\textbf{g}_{k}^{H})\big) \odot\big(\textbf{G}(\textbf{F}_{a}+\textbf{F}_{b})\textbf{G}_{k}^{H}\big)^{T},
\end{flalign}	
 \begin{flalign}
 \textbf{s}_{k} =\frac{2}{\pi}\text{diag}\big(\textbf{h}_{k}^{H}(\textbf{F}_{a}+\textbf{F}_{b})\textbf{G}^{H}\big)\textbf{g}_{k}.
\end{flalign}	
At this stage, the objective function (\ref{objective function P3.1}) can be reformulated as
 \begin{flalign}
f_{R3}(\overline{\mathbf{V}}_{t},\overline{\mathbf{V}}_{r})& =
\sum_{k=0}^{k_{0}}\Big(2b_{k}\sqrt{(1+a_{k})\text{Tr}(\overline{\textbf{Q}}_{k}\overline{\textbf{V}}_{t})}-b_{k}^{2}\text{Tr}(\overline{\textbf{S}}_{k}\overline{\textbf{V}}_{t}) \Big) \notag \\
+\sum_{k_{0}+1}^{K}&\Big(2b_{k}\sqrt{(1+a_{k})\text{Tr}(\overline{\textbf{Q}}_{k}\overline{\textbf{V}}_{r})}-b_{k}^{2}\text{Tr}(\overline{\textbf{S}}_{k}\overline{\textbf{V}}_{r}) \Big).
\end{flalign}	

And the problem ($P_{3}$) can be rewritten as
\begin{subequations}
	\begin{flalign}
		(\overline{P}_{3}): &\underset{\overline{\mathbf{V}}_{t}\overline{\mathbf{V}}_{r}}{\text{max}}f_{R3}(\overline{\mathbf{V}}_{t},\overline{\mathbf{V}}_{r}),\label{objective function P3.2} \\ 
		\text{s.t.} \ \ \ &   [\overline{\textbf{V}}_{t}]_{n,n}+[\overline{\textbf{V}}_{r}]_{n,n}=1,n=1,\,.\,.\,.\,,\,N, \label{constraint1 for P3} \\
   &[\overline{\textbf{V}}_{t}]_{N+1}=[\overline{\textbf{V}}_{r}]_{N+1}=1, \label{constraint2 for P3}\\
   &\overline{\textbf{V}}_{t}\succeq \textbf{0},\overline{\textbf{V}}_{r}\succeq \textbf{0}, \label{constraint3 for P3}\\
&\text{Rank}(\overline{\textbf{V}}_{t})=\text{Rank}(\overline{\textbf{V}}_{r})=1, \label{sdr rank constraint 2} \\
&\sum_{k=1}^{K}c_{k}\leq \widetilde{R}_{c,k} ,\forall k.  \label{constraint5 for P3}
\end{flalign}
\end{subequations}
Here, the constraint (\ref{constraint5 for P3}) can be expressed as
\begin{flalign}
\Big(\text{Tr}(\overline{\textbf{S}}_{k}\overline{\textbf{V}}_{t})+\frac{1}{\gamma}\Big)\text{2}^{\sum_{k=1}^{K}c_{k}} \leq &\text{Tr}(\overline{\textbf{S}}_{k}\overline{\textbf{V}}_{t})+\notag\\& \text{Tr}(\overline{\textbf{W}}_{k}\overline{\textbf{V}}_{t})+\frac{1}{\gamma},{k\in\mathcal{T},} \label{transferred constraint 5 for P3_1}
\end{flalign}
\begin{flalign}
	\Big(\text{Tr}(\overline{\textbf{S}}_{k}\overline{\textbf{V}}_{r})+\frac{1}{\gamma}\Big)\text{2}^{\sum_{k=1}^{K}c_{k}} \leq &\text{Tr}(\overline{\textbf{S}}_{k}\overline{\textbf{V}}_{r})+\notag\\&\text{Tr}(\overline{\textbf{W}}_{k}\overline{\textbf{V}}_{r})+\frac{1}{\gamma},{k\in\mathcal{R}}, \label{transferred constraint 5 for P3_2}
\end{flalign}
where
\begin{flalign}
		\overline{\textbf{W}}_{k}=\begin{bmatrix}
		\textbf{W}_{k} & \textbf{w}_{k} \\
		\textbf{w}_{k}^{H} & \textbf{f}_{c}^{H}\textbf{h}_{k}\textbf{h}_{k}^{H}\textbf{f}_{c}
	\end{bmatrix},
\end{flalign}
with
\begin{flalign}
 \textbf{W}_{k}=\big(\text{diag}(\textbf{g}_{k})\textbf{V}_{x}\text{diag}(\textbf{g}_{k}^{H})\big) \odot\big(\textbf{G}\textbf{f}_{c}\textbf{f}_{c}^{H}\textbf{G}^{H}\big)^{T},
\end{flalign}

\begin{flalign}
\textbf{w}_{k} =\frac{2}{\pi}\text{diag}(\textbf{h}_{k}^{H}\textbf{f}_{c}\textbf{f}_{c}^{H}\textbf{G}^{H})\textbf{g}_{k}.
\end{flalign}
To this end, apart from the non-convex rank-one constraint (\ref{sdr rank constraint 2}), all other constraints are convex. To solve it, we resort to the PF method to handle constraint (\ref{sdr rank constraint 2})\cite{Robust Secrecy-Energy Efficient Beamforming for Jittering UAV}. Noticing the fact:
\begin{flalign}
\text{Tr}(\overline{\textbf{V}}_{i})-\lambda_{max}(\overline{\textbf{V}}_{i}) = 0 \Rightarrow \text{Rank}(\overline{\textbf{V}}_{i}) = 1, \ {i=\{v,r\}},
\end{flalign}
where $\lambda_{max}(\overline{\textbf{V}}_{i})$ denotes the largest eigenvalue of $\overline{\textbf{V}}_{i}$, we incorporate
the term $\text{Tr}(\overline{\textbf{V}}_{i})-\lambda_{max}(\overline{\textbf{V}}_{i})$ into (\ref{objective function P3.2}) as the penalty term and reformulate he objective function as
\begin{flalign}
f_{R3}(\overline{\mathbf{V}}_{t},\overline{\mathbf{V}}_{r})-\eta \sum \nolimits_{{i=\{v,r\}}} \Big(\text{Tr}(\overline{\textbf{V}}_{i})
-\lambda_{max}(\overline{\textbf{V}}_{i})\Big),
\end{flalign}
where $\eta$ is the penalty coefficient. Following the idea of PF, the new objective function aims to maximize $f_{R3}(\overline{\mathbf{V}}_{t},\overline{\mathbf{V}}_{r})$ and minimize the penalty term
$\sum \nolimits_{{i=\{v,r\}}}\Big(\text{Tr}(\overline{\textbf{V}}_{i})-\lambda_{max}(\overline{\textbf{V}}_{i})\Big)$, simultaneously. When the penalty term approaches zero, $\textbf{V}_{i}$
can be considered to have only one non-zero eigenvalue, and thus the rank-one constraint (\ref{sdr rank constraint 2}) is satisfied. Then, by applying SCA method, using the first-order Taylor expansion, the convex term $\lambda_{max}(\overline{\textbf{V}}_{i})$ should satisfy the following inequality
\begin{flalign}
\lambda_{max}(\overline{\textbf{V}}_{i})\geq \lambda_{max}(\overline{\textbf{V}}_{i}^{n}) + \text{Tr}\Big(\mathbf{\iota}_{max}(\overline{\textbf{V}}_{i}^{n}) \mathbf{\iota}_{max}^{H}(\overline{\textbf{V}}_{i}^{n})(\overline{\textbf{V}}_{i}-\overline{\textbf{V}}_{i}^{n})\Big),
\end{flalign}
where $\overline{\textbf{V}}_{i}^{n}$ is a feasible point of $\overline{\textbf{V}}_{i}$ and $\mathbf{\iota}_{max}(\overline{\textbf{V}}_{i}^{n})$ represents the corresponding unit-norm eigenvector of $\lambda_{max}(\overline{\textbf{V}}_{i}^{n})$.
Thus, the objective function can be further reformulated as
\begin{flalign}
f_{R3}'(\overline{\mathbf{V}}_{t},\overline{\mathbf{V}}_{r}) = f_{R3}(\overline{\mathbf{V}}_{t},\overline{\mathbf{V}}_{r})-\eta \sum \nolimits_{{i=\{v,r\}}} \Big(\text{Tr}(\overline{\textbf{V}}_{i})
- \notag \\ \lambda_{max}(\overline{\textbf{V}}_{i}^{n}) - \text{Tr}\Big(\mathbf{\iota}_{max}(\overline{\textbf{V}}_{i}^{n}) \mathbf{\iota}_{max}^{H}(\overline{\textbf{V}}_{i}^{n})(\overline{\textbf{V}}_{i}-\overline{\textbf{V}}_{i}^{n})\Big)\Big).
\end{flalign}
At this stage, $(\overline{P}_{3})$ can be rewritten as
\begin{subequations}
	\begin{flalign}
		(\widehat{P_{3}}): \ & \underset{\overline{\mathbf{V}}_{t}\overline{\mathbf{V}}_{r}}{\text{max}} \ \  f_{R3}'(\overline{\mathbf{V}}_{t},\overline{\mathbf{V}}_{r}), \\
		\text{s.t.} \ \ \ \ &   \text{(\ref{constraint1 for P3})}, \text{(\ref{constraint2 for P3})}, \text{(\ref{constraint3 for P3})}, \text{(\ref{transferred constraint 5 for P3_1})},
		\text{(\ref{transferred constraint 5 for P3_2}) }. \notag
	\end{flalign}
\end{subequations}
Till now, the problem can be solved by using the CVX tool.

\subsection{Complexity and Optimality Analysis}
\emph{1) Complexity Analysis}: The computational complexity represents the number of multiplication operations in an expression. Generally, the highest-order term is taken while lower-order terms are ignored. Hence, for updating auxiliary variables $a_{k}^{\ast}$ and $b_{k}^{\ast}$ is $\mathcal{O}(2KM^{3})$. For solving sub-problem $(P_{2})$, we employ the SCA method, which contributes $\mathcal{O}(M^{3})$ complexity per iteration. Assuming $l_{1}$ iterations for SCA, the total complexity of $(P_{2})$ becomes $\mathcal{O}(l_{1}(K+1)M^{3.5})$. Regarding sub-problem $(P_{3})$, assume $l_{2}$ iterations for SCA, the computational complexity becomes $\mathcal{O}(2l_{2}(N+1)^{3.5})$.

\emph{2) Optimality Analysis}: Based on the AO framework, the auxiliary variables $a$, $b$ and sub-problems $(P_{2})$, $(P_{3})$ are solved in an iterative way. For $a$ and $b$, the optimal solution can be found by taking partial derivative while for $(P_{2})$ and $(P_{3})$, both of which yield locally optimal solutions. Hence, the proposed FP-AO algorithm for solving $(P_{0})$ guarantees convergence to at least a locally optimal solution, as substantiated by the proof presented in \cite{A unified algorithmic framework for block-structured}.

\section{Numerical Results}
In this section, associated numerical results are provided to evaluate the performance of the proposed STAR-RIS aided RSMA communication system and to  validate the benefits of the robust beamforming design considering HWI.
\begin{table}
	\begin{center}
		\caption{Simulation parameters.}
		\label{tab1}
		\renewcommand{\arraystretch}{1.2} 
		\begin{tabular}{| c | c | c |}
			\hline
			\makecell{Parameters} & Description & value\\
			
			\hline
			$M$& the amount of antennas & 4\\
			\hline
			$K$&the amount of users &4\\ 
			\hline
			\makecell{$\sigma^{2}$}& noise power & -70 dBm\cite{Robust and Secure Transmission Design for }\\
			\hline
			$\varepsilon$& rician factor & 10\\
			\hline 
			$L_{0}$& path loss at 1m & -30 dB\\
			\hline
			$c_{G}$& path loss component of G & 2.6\\
			\hline 
			$c_{g,k}$& path loss component of g & 2.2\\
			\hline
			$c_{h}$&path loss component of h & 5\\
			\hline
			$\epsilon$&convergence accuracy& $10^{-4}$\\
			\hline
		\end{tabular}
	\end{center}
\end{table}
\begin{figure}[!t]
	\centering
	\includegraphics[width=3.3in]{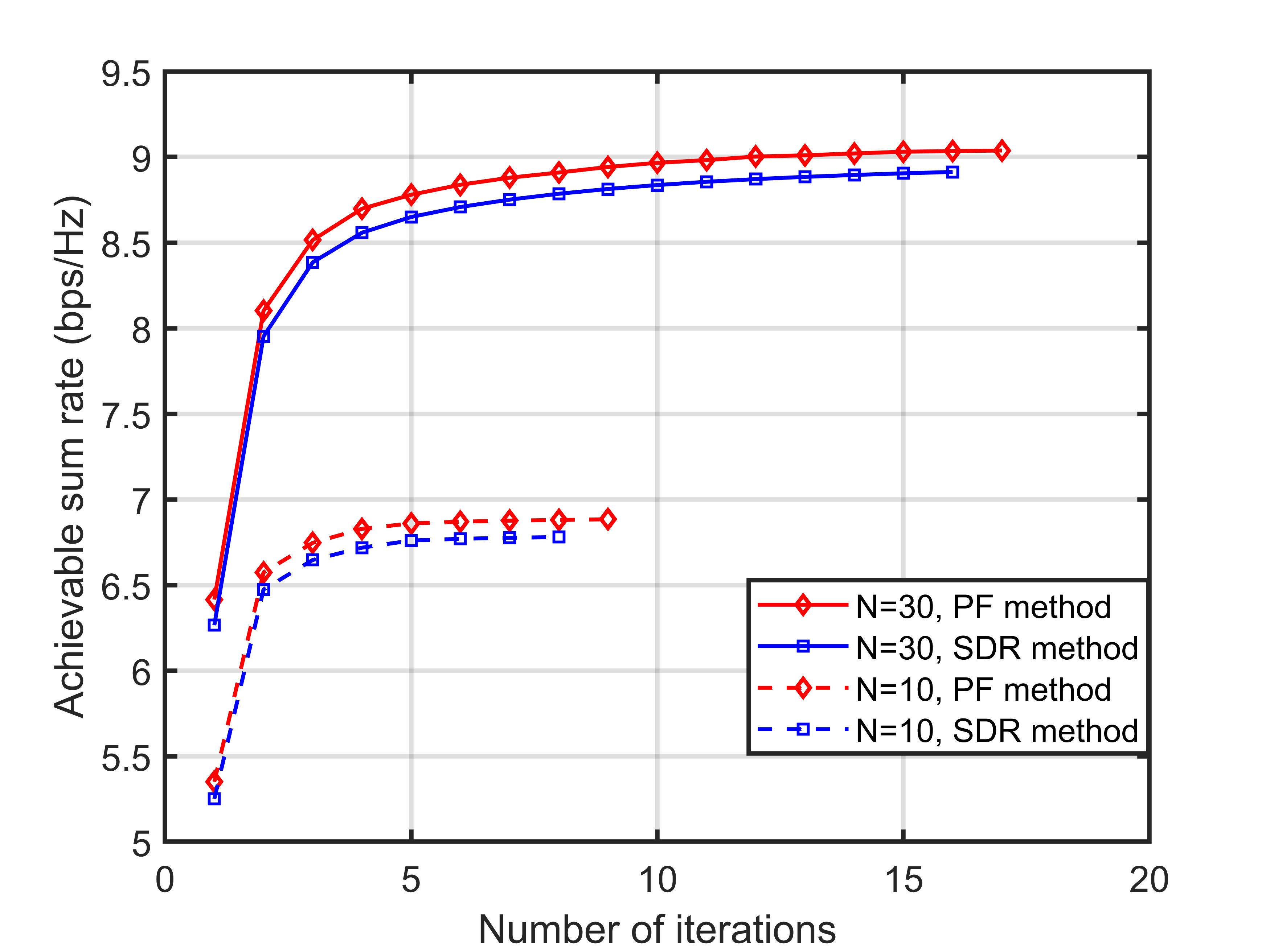}
	\caption{Achievable sum rate versus the number of iterations to get convergence under different number of STAR-RIS elements $N$ and different algorithms where $\mu_{t}=\mu_{r}=0.01$, $P_{max}$ = 5 W.}
	\label{fig_2}
\end{figure}
\subsection{Simulation Setup}
In the simulations, we consider  a single-cell situation that includes one BS, multiple users and one STAR-RIS which is installed close to the users. In a two-dimensional (2D) coordinate the BS is located at (0,0)m, STAR-RIS at (40,0)m, four users are introduced and each two of them are randomly distributed in two circular regions centered at (-2, 40)m and (2, 40)m with a radius 2 m, corresponding to the transmission region and reflection region respectively. The simulation parameters and descriptions are shown in Table I.

We consider that the links between the BS and the STAR-RIS, as well as those between the STAR-RIS and users, are characterized by LoS channels. Thus, both $\textbf{G}$ and $\textbf{g}_{k}$ are assumed to follow the Rician distribution.
	\begin{flalign}
		&\textbf{G} = \beta_{G}\bigg(\sqrt{\frac{\varepsilon}{1+\varepsilon}}\text{\textbf{G}}^{\text{LoS}}+(\sqrt{\frac{1}{1+\varepsilon}}\text{\textbf{G}}^{\text{NLoS}}\bigg), \\
			&\textbf{g}_{k} =\beta_{g,k}\bigg(\sqrt{\frac{\varepsilon}{1+\varepsilon}}\text{\textbf{g}}_{k}^{\text{LoS}}+(\sqrt{\frac{1}{1+\varepsilon}}\text{\textbf{g}}_{k}^{\text{NLoS}}\bigg),
	\end{flalign}	
$\varepsilon$ represents the Rician factor, $\beta_{G}$ and $\beta_{g,k}$ denote the pathloss from the BS to the STAR-RIS and from the STAR-RIS to the $k$-th user, respectively. $\text{\textbf{G}}^{\text{LoS}}$ and $\text{\textbf{g}}_{k}^{\text{LoS}}$ are the corresponding LoS components, while $\text{\textbf{G}}^{\text{NLoS}}$ and $\text{\textbf{g}}_{k}^{\text{NLoS}}$ are the non-LoS (NLoS) components \cite{Robust and Secure Transmission Design for }. Here, the pathloss $\beta_{G}$ and $\beta_{g,k}$ are expressed as $\sqrt{L_{0}d_{G}^{-c_{G}}}$, $\sqrt{L_{0}d_{g,k}^{-c_{g,k}}}$ where $d$ is the distance between the two objects. $c$ represents the corresponding pathloss component, and $L_{0}$ is the pathloss at 1m. Assume the channel between BS to the users follow the same model but without LoS component. 
\begin{figure}[!t]
	\centering
	\includegraphics[width=3.3in]{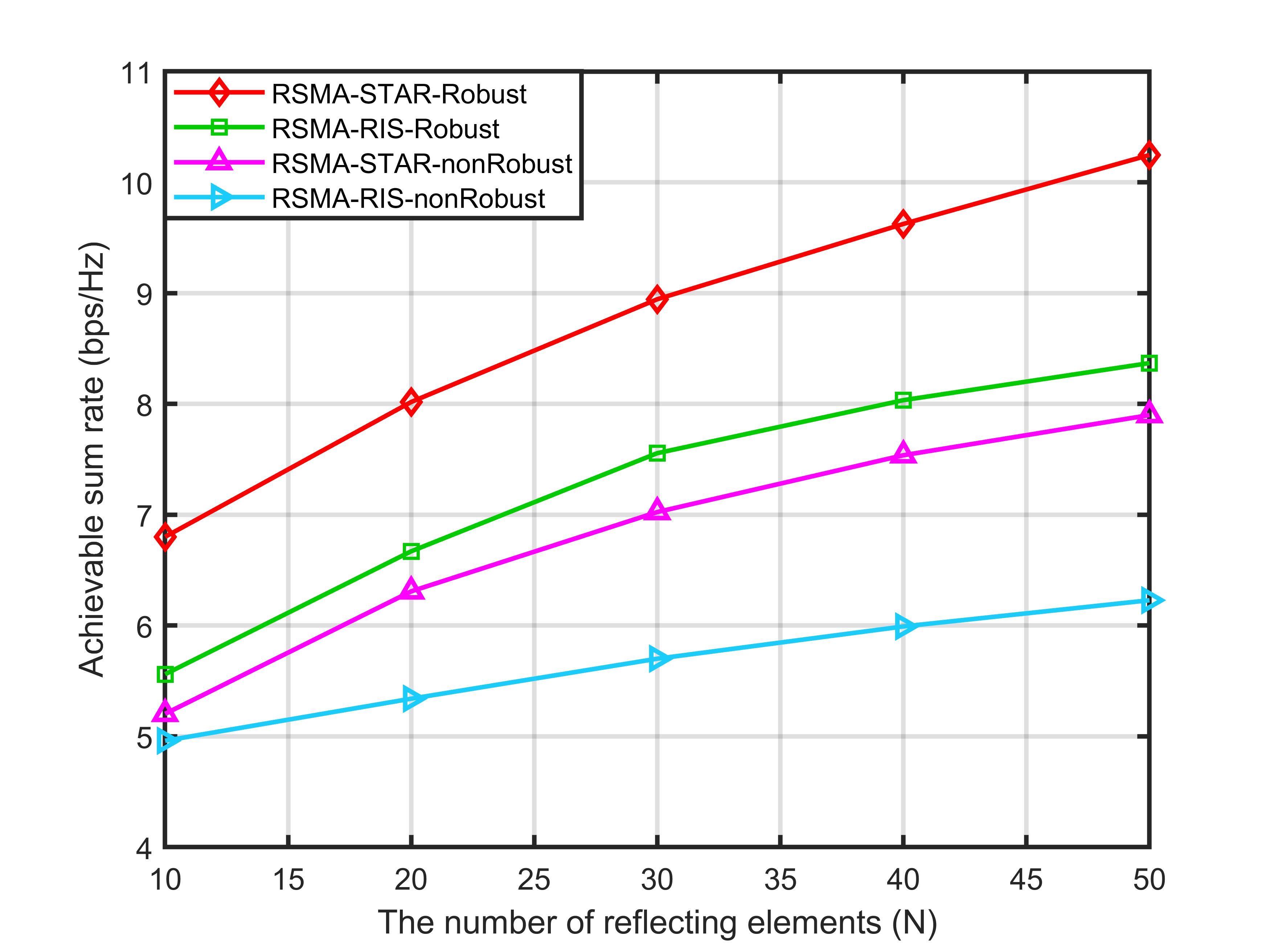}
	\caption{Achievable sum rate versus the number of reflecting elements $N$ under different type of RIS and beamforming design where $P_{max}$ = 5 W, $\mu_{t}=\mu_{r}=0.01$.}
	\label{fig_2}
\end{figure}

\begin{figure}[!t]
	\centering
	\includegraphics[width=3.3in]{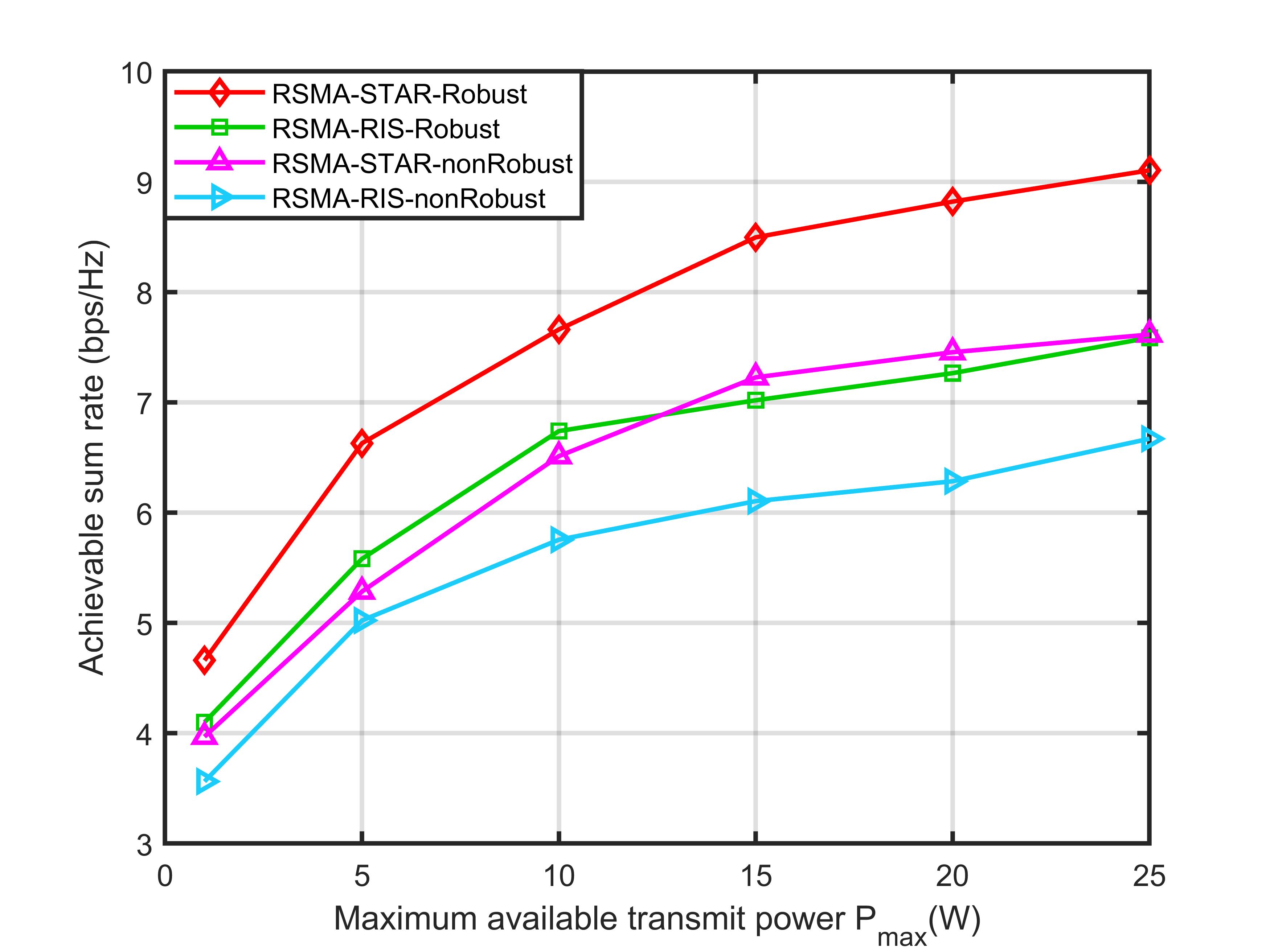}
	\caption{Achievable sum rate versus the transmit power $P_{max}$ under different type of RIS and beamforming design where $N$ = 10, $\mu_{t}=\mu_{r}=0.01$.}
	\label{fig_2}
\end{figure}
We consider several schemes as baseline for comparison. (1) RSMA-STAR-nonRobust scheme utilizes RSMA and STAR-RIS in the system but neglects the influencing factors of HWI in the algorithm design; (2) RSMA-RIS-Robust scheme also employs RSMA method and considering the influence of HWI, while the STAR-RIS operates in a conventional RIS mode, where $N/2$ elements are used to transmit signals only, and $N/2$ elements are used to reflect signals only to facilitate full-space coverage; (3) RSMA-RIS-nonRobust scheme employs RSMA and conventional RIS mode, without considering HWI; (4) NOMA-STAR-Robust scheme adopts NOMA multiple-access technique aided by STAR-RIS while accounting for the presence of HWI where the deployment of NOMA scheme follows the one in \cite{Achievable Rate Analysis of the STAR-RIS}; (5) OMA-STAR-Robust scheme adopts OMA multiple-access technique aided by STAR-RIS while accounting for the presence of HWI. Here, we utilizes orthogonal frequency division multiple access (OFDMA) as the scheme of OMA which divided the frequency band equally into $K$ parts for $K$ users, respectively.

\subsection{Simulation Result}
\emph{1) Convergence Performance}: Fig. 2 demonstrates that the proposed algorithm consistently converges to stable solutions through successive iterations, across varying numbers of STAR-RIS elements and users. As the number of STAR-RIS elements increases, it can be observed that the system's achievable sum rate exhibits improved performance. This enhancement is attributed to the fact that, under the same energy constraints, a larger number of elements offers greater degrees of freedom (DoF) for electromagnetic wave manipulation, leading to more precise beamforming and more flexible transmission strategy. However, it is evident that increasing the number of $N$ also adds complexity to the problem, resulting in a higher number of iterations required for convergence. In the figure, it also compares the performance of different algorithm models by replacing the PF algorithm when solving the rank-1 constraint in ($\overline{P}_{3}$) with the semidefinite relaxation (SDR) algorithm. The results demonstrate that our proposed algorithm outperforms the SDR-based approach. A possible explanation lies in the performance loss caused by the Gaussian randomization used in SDR.

\emph{2) Gains of STAR-RIS and the Robust Beamforming Design}:
To highlight the performance gains enabled by STAR-RIS and robust beamforming, Fig. 3 compares robust and non-robust designs in systems employing STAR-RIS and conventional passive RIS. It is observed that STAR-RIS consistently outperforms conventional RIS, regardless of whether HWI is taken into account. The primary reason for this improvement lies in the fact that STAR-RIS can fully exploit the available DoF to control beam propagation. In contrast, in conventional passive RIS-aided networks, the performance gain is limited due to the optimization being restricted solely to the reflection phases of the elements. Furthermore, when the same type of RIS is employed, the robust beamforming design consistently outperforms the non-robust approach. As the number of $N$ increases, the performance advantage of the robust scheme over the non-robust scheme becomes increasingly evident. When $N=10$, the achievable sum rate of RSMA-STAR-Robust scheme is approximately 21\% higher than that of RSMA-STAR-nonRobust. When $N$ reaches 50, this gap widens to 27.5\%. This demonstrates the exceptional capability of the proposed FP-AO algorithm in compensating for HWI.

Fig. 4 plots the maximum system power budget versus the achievable sum rate. It shows that when $P_{max}$ is small, the system achievable sum rate increases rapidly with the growth of the transmit power. As $P_{max}$ rises from 1W to 5W, all four schemes achieve a significant performance improvement of approximately 40\%. However, this trend gradually stabilizes as $P_{max}$ increases. When $P_{max}$ reaches 15W, further increasing the transmit power provides only a very marginal gain to the system. This stability appears because the distortion noise induced by HWI is proportional to the transceiver signal power, thereby establishing an upper bound for the performance of the system. Under the same power budget, it is evident that the robust design of the STAR-RIS aided RSMA system we proposed has better performance than the other three, demonstrates the necessity of considering HWI in the algorithm design.

\begin{figure}[!t]
	\centering
	\includegraphics[width=3.3in]{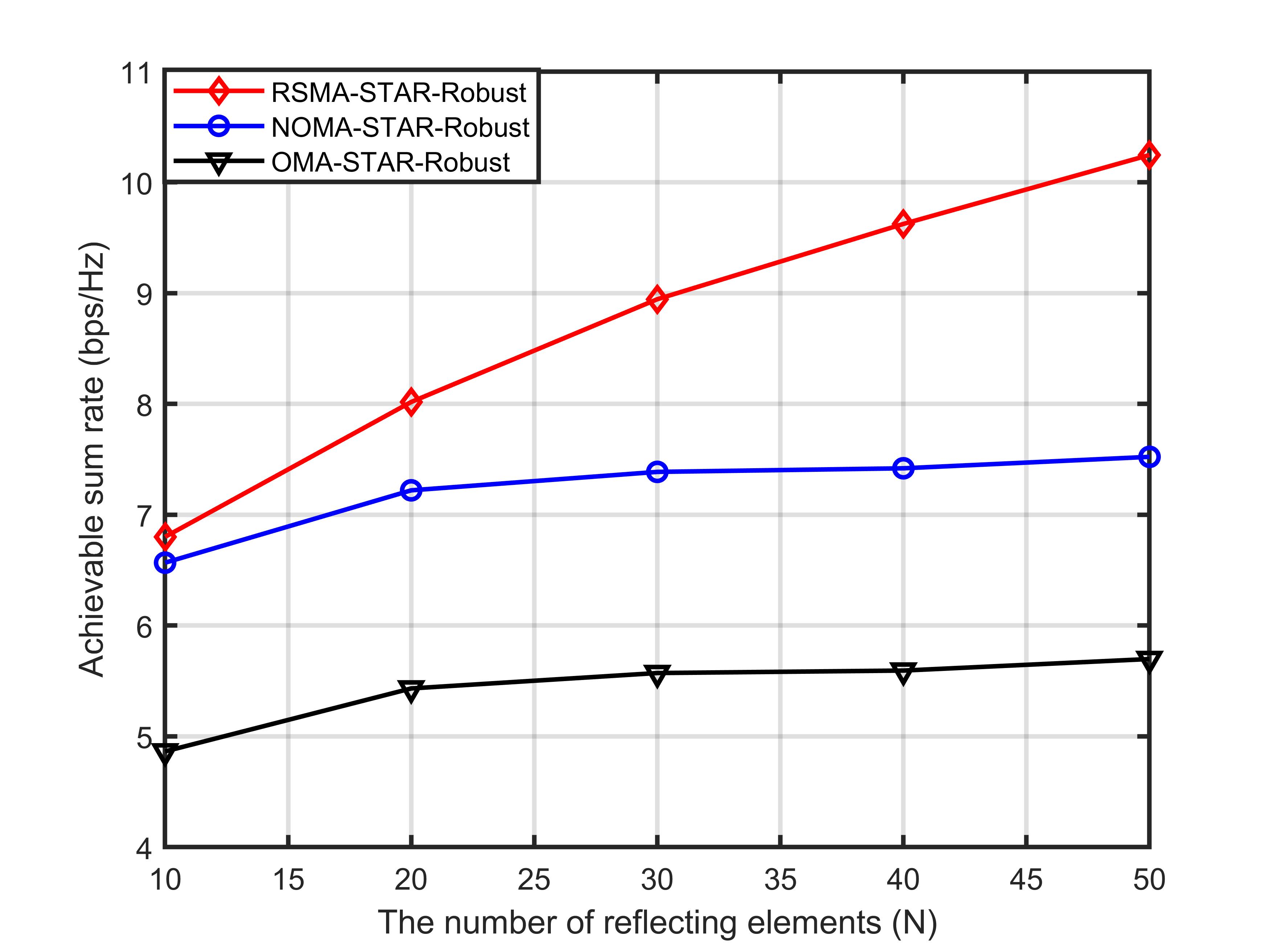}
	\caption{Achievable sum rate versus the number of STAR-RIS elements $N$ under different multiple access schemes where $P_{max}$ = 5 W, $\mu_{t}=\mu_{r}=0.01$.$\ $ }
	\label{fig_2}
\end{figure}

\begin{figure}[!t]
	\centering
	\includegraphics[width=3.3in]{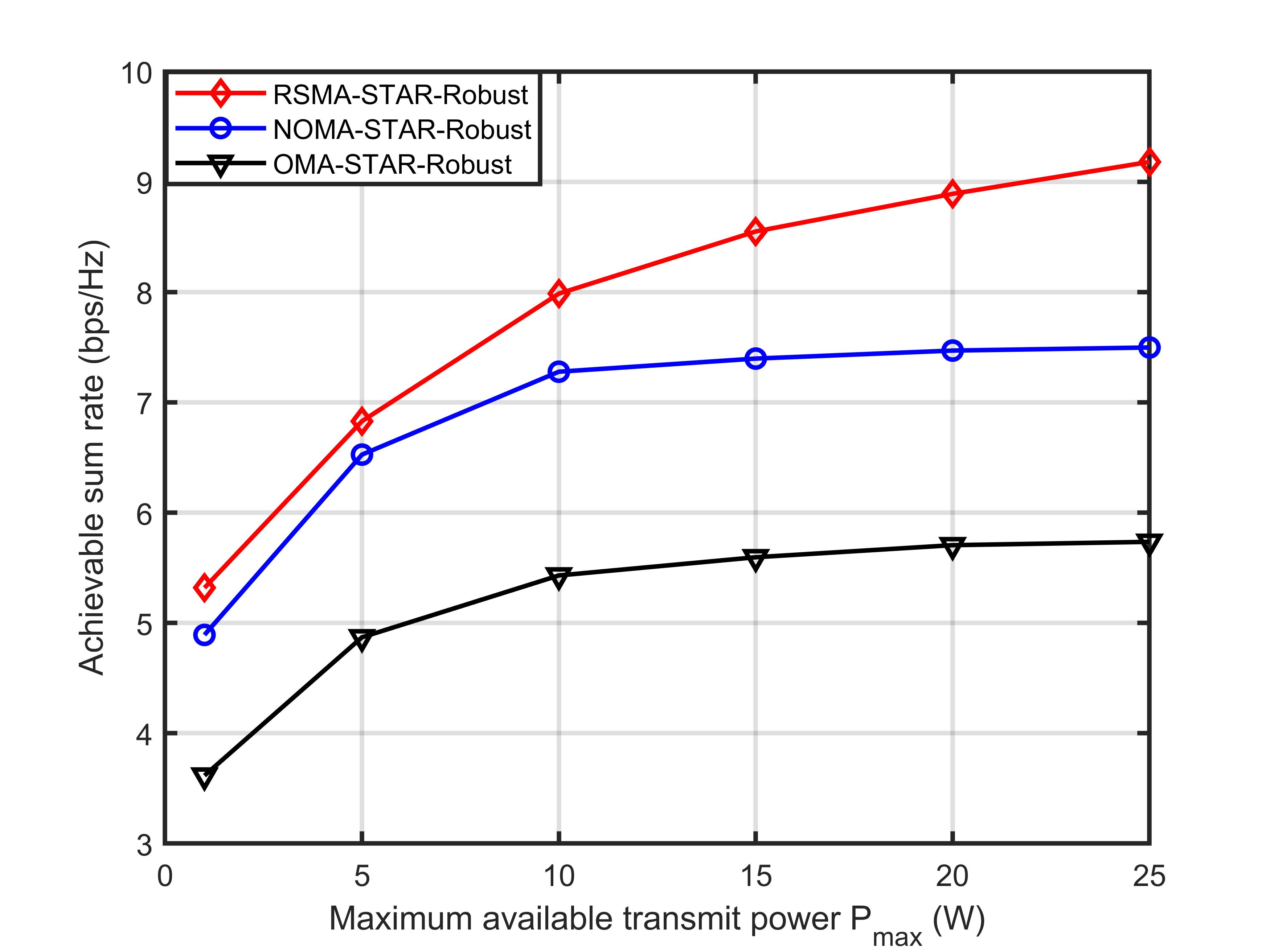}
	\caption{Achievable sum rate versus the transmit power $P_{max}$ under different  multiple access schemes where $N$ = 10, $\mu_{t}=\mu_{r}=0.01$.}
	\label{fig_2}
\end{figure}
\emph{3) Effect of RSMA Scheme on the Network}: Fig. 5 presents the system achievable sum rate versus the number of reflecting elements $N$. As 
$N$ increases, the performance of OMA and NOMA improves only marginally and even saturates when $N$ becomes sufficiently large, due to the influence of HWI on reflecting elements. RSMA shows a more stable and superior trend which achieved nearly 50\% performance gain as $N$ increased from 10 to 50. At the point of $N$ = 50,  RSMA scheme delivers a 36\% improvement over NOMA and a 75\% advantage compared to traditional OMA scheme. This behavior suggests that RSMA is more resilient to the HWI on the reflecting elements, making it a more robust choice for realistic STAR-RIS systems. Fig. 6 compares the performance of different multiple access schemes in STAR-RIS Robust system. The results show that RSMA achieves the best performance across all transmit power levels, and its advantage becomes more significant as $P_{max}$ increases. When power level is low, its increasing leads to substantial gains in system performance. However, as $P_{max}$ grows further, the performance improvements of all three schemes progressively decreases. This diminishing effect is likely caused by signal distortion under high-power conditions due to HWI's existing.

\emph{4) Impact of the Transceiver HWI Coefficients}\footnote{According to \cite{Massive MIMO Systems With Non-Ideal}, the choice of HWI coefficients is closely related to the requirement of error vector magnitude (EVM) specified in the 3GPP standard. For NR and LTE systems, the HWI coefficients should be less than or equal to $0.175^{2}$.}: Fig. 7 shows the achievable sum rate of the system users under different transmission schemes versus the HWI coefficients.  It is apparently noted that as the hardware impairments of the transceivers become more severe, the performance of each system declines accordingly, due to signal distortion caused by hardware degradation. It is observed that whether the robust design is implemented or not, the STAR-RIS aided RSMA system always performs better than the conventional RIS. Under low HWI conditions, the RSMA-STAR-Robust scheme achieves approximately a 23\% gain in achievable sum rate compared to the RSMA-RIS-Robust scheme, while the RSMA-STAR-nonRobust scheme shows a 12\% improvement over RSMA-RIS-nonRobust under high HWI conditions. Moreover, the robust design we proposed significantly outperforms non-robust designs under various levels of HWI impact and different type of RIS deployment, demonstrating the necessity of considering HWI in beamforming algorithms to overcome the impact of signal distortion interference and to better unlock the capabilities of RSMA scheme.

\section{Conclusion}
In this work, we explored a downlink RSMA communication system aided by a STAR-RIS, where the HWI is considered at both the transceivers and the STAR-RIS side. An achievable sum rate maximization problem is formulated by jointly optimizing the BS beamforming vectors for both the common and private streams, the STAR-RIS coefficients, and the actual transmission rate of the common stream. To tackle this highly coupled non-convexity problem, we first employ FP method to reformulate the objective function into a more tractable form, then solve the resulting sub-problems within an AO framework by leveraging SCA and PF methods. Simulation results show that the proposed FP-AO algorithm converges efficiently and that the RSMA-STAR-Robust scheme significantly outperforms benchmark schemes in terms of the achievable sum rate.

\begin{figure}[!t]
	\centering
	\includegraphics[width=3.3in]{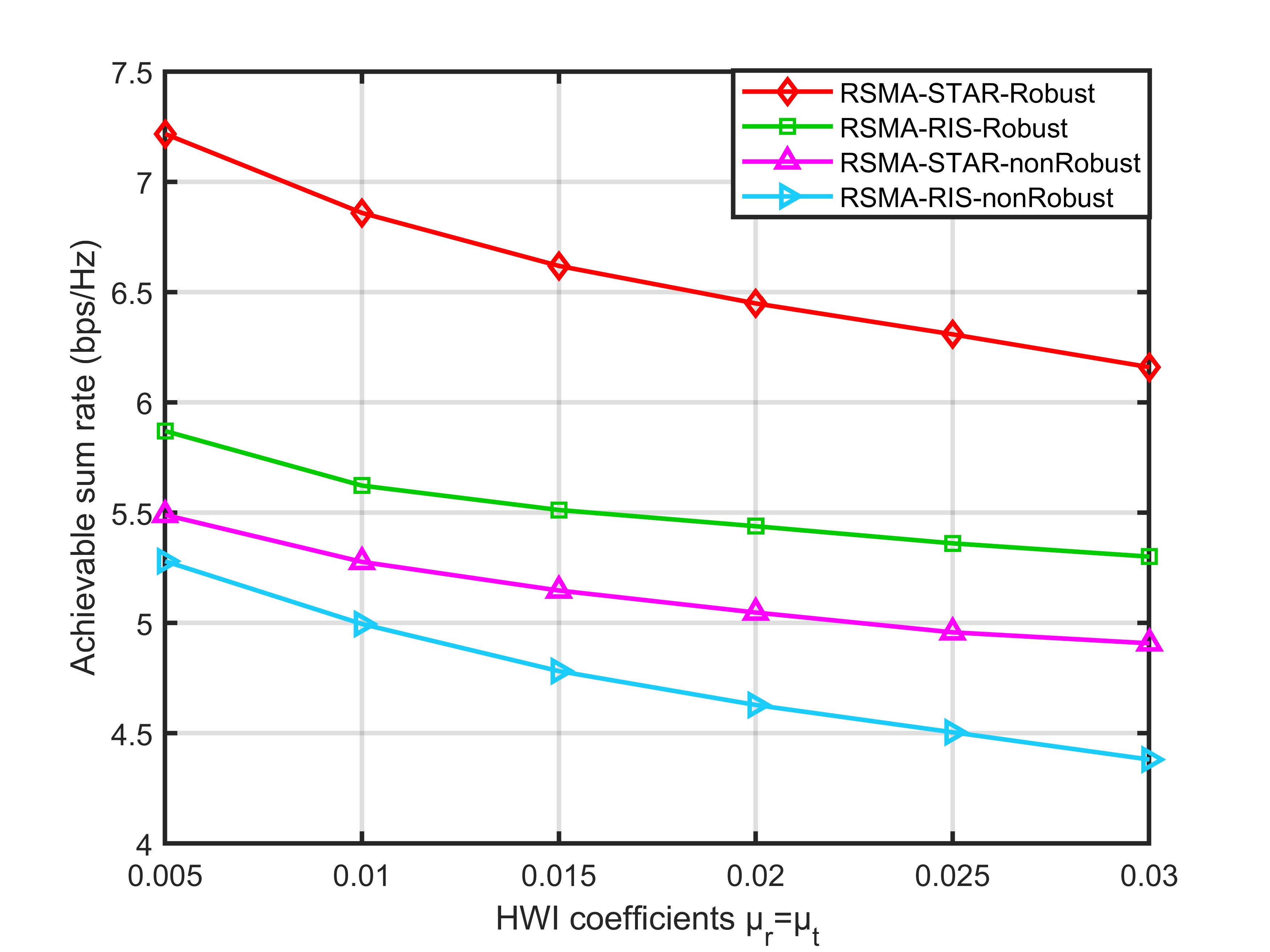}
	\caption{Achievable sum rate versus the HWI coefficients $\mu_{t}$,$\mu_{r}$ under different type of RIS and beamforming design where $N$ = 10, $P_{max}$ = 5 W.}
	\label{fig_2}
\end{figure}

\end{document}